\documentclass[twocolumn, resetfootnote]{aastex7}
\usepackage{tabularx}
\newcolumntype{Y}{>{\centering\arraybackslash}X}
\usepackage{tabularray}
\usepackage{threeparttable}
\usepackage{booktabs}
\UseTblrLibrary{booktabs}
\usepackage[table]{xcolor}
\usepackage{multirow}

\usepackage[autostyle, english = american]{csquotes}
\MakeOuterQuote{"}
\usepackage{mhchem}
\usepackage{float}
\usepackage{rotating}
\renewcommand{\thempfootnote}{\arabic{mpfootnote}}

\begin{document}

\title{A Multi-Wavelength Study of Comet C/2022 E3 (ZTF): Complementary ALMA and JWST Investigations of {Water and Methanol} in Cometary Comae}

\author[0000-0001-6192-3181]{K. D. Foster}
\affiliation{Department of Chemistry, University of Virginia, Charlottesville, VA 22904, USA}
\email{kfoster@virginia.edu}

\author[0000-0001-8233-2436]{M. A. Cordiner}
\affiliation{Solar System Exploration Division, Astrochemistry Laboratory Code 691, NASA Goddard Space Flight Center, 8800 Greenbelt Rd, Greenbelt, MD 20771, USA}
\affiliation{Department of Physics, Catholic University of America, Washington, D.C. 20064, USA}
\email{martin.cordiner@nasa.gov}

\author[0000-0002-6006-9574]{Nathan X. Roth}
\affiliation{Solar System Exploration Division, Astrochemistry Laboratory Code 691, NASA Goddard Space Flight Center, 8800 Greenbelt Rd, Greenbelt, MD 20771, USA}
\affiliation{Department of Physics, American University, 4400 Massachusetts Ave NW, Washington, DC 20016, USA}
\email{nathaniel.x.roth@nasa.gov}

\author[0000-0001-7694-4129]{S. N. Milam}
\affiliation{Solar System Exploration Division, Astrochemistry Laboratory Code 691, NASA Goddard Space Flight Center, 8800 Greenbelt Rd, Greenbelt, MD 20771, USA}
\email{stefanie.n.milam@nasa.gov}

\author[0000-0001-9479-9287]{A. J. Remijan}
\affiliation{National Radio Astronomy Observatory, Charlottesville, VA 22093, USA}
\email{aremijan@nrao.edu}

\author[0000-0003-2414-5370]{N. Biver}
\affiliation{LIRA, Observatoire de Paris, Université PSL, CNRS, Sorbonne Université, Université Paris Cité, CY Cergy Paris Université, 5 place Jules Janssen, 92190 Meudon, France}
\email{Nicolas.Biver@obspm.fr}

\author[0000-0002-1545-2136]{J. Boissier}
\affiliation{Institut de Radioastronomie Millimetrique, 300 rue de la Piscine, F-38406, Saint Martin d’Heres, France}
\email{boissier@iram.fr}

\author{J. Crovisier}
\affiliation{LIRA, Observatoire de Paris, Université PSL, CNRS, Sorbonne Université, Université Paris Cité, CY Cergy Paris Université, 5 place Jules Janssen, 92190 Meudon, France}
\email{Jacques.Crovisier@obspm.fr}

\author[0000-0002-4336-0730]{Y.-J. Kuan}
\affiliation{Center of Astronomy and Gravitation, and Department of Earth Sciences, National Taiwan Normal University, Taipei, Taiwan, ROC}
\affiliation{Institute of Astronomy and Astrophysics, Academia Sinica, Taipei, Taiwan, ROC}
\email{kuan@ntnu.edu.tw}

\author[0000-0002-0500-4700]{D. C. Lis}
\affiliation{Jet Propulsion Laboratory, California Institute of Technology, 4800 Oak Grove Drive, Pasadena, CA 91109, USA}
\email{dariusz.c.lis@jpl.nasa.gov}

\begin{abstract}
    Long-period comets, which are often considered to be representative of material in the protoplanetary disk that formed the Solar System, are ideal to investigate the question of chemical inheritance in astronomy. Determining the chemistry of comets, both individually and as a population, has become of great importance in comparative studies against sources representative of evolutionary precursors to planetary systems. Contemporaneous observations of long-period comet C/2022 E3 (ZTF) were obtained with the JWST and the Atacama Large Millimeter/submillimeter Array (ALMA) in early 2023 March. {This work focuses on \ce{CH3OH} measurements from both ALMA and JWST as well as \ce{H2O} measurements from JWST.} Radiative transfer modeling of \ce{CH3OH} and \ce{H2O} was performed to investigate spatial variations in rotational temperature, column density, and production rates, as well as a comparison of derived values between the two telescopes. Most of the spatial distributions of the modeled values are centrally peaked, and the modeled values from JWST are all within the error bars of the average values from ALMA. C/2022 E3 (ZTF) also displays an enhancement in modeled rotational temperature in the anti-Sunward direction that is shown to be statistically significant. Based on non-LTE radiative transfer modeling, the declining \ce{H2O} rotational temperatures as a function of nucleocentric distance observed by JWST can be explained primarily as a result of rotational line cooling. The values derived in this work are in general agreement with single-dish millimeter-wave observations.
\end{abstract}


\section{INTRODUCTION} \label{introduction}

Since comets are assumed to have formed from the same nebula as the rest of the Solar System, their compositions may be indicative of the chemical complexity of the protoplanetary disk and prestellar core phases \citep{Mumma2011, Dones2015}. Comets are typically characterized based on their orbit and source reservoir as either an elliptic comet (EC) from the Kuiper Belt or scattered disk, or a nearly isotropic comet (NIC) from the Oort Cloud \citep{Mumma2011}. As ECs have shorter orbital periods, they experience more solar radiation and as a result may be more processed than NICs \citep{Weissman2020}. Thus, NICs are suitable targets to investigate whether the chemical complexity in the precursor protoplanetary disk stage is carried through to the planetesimal material that formed the comets, a concept known as inheritance. The inheritance of chemical complexity in a source from its precursor source is not specific to the transition from a protoplanetary disk to a planetary system and the extent of the complexity inherited{\textemdash}from all (inheritance scenario), to some (partial reset scenario), or to none (full reset scenario){\textemdash}is an ongoing subject of debate \citep{Ehrenfreund2000, Caselli2012, Drozdovskaya2019}. Previous works have compared ratios of molecular abundances or isotopic fractionation between different types of sources \citep{Caselli2012, Drozdovskaya2019}. For example, there have been investigations of molecules containing carbon, hydrogen, nitrogen, oxygen, phosphorus, and sulfur (CHNOPS) where the molecular abundances of these molecules in comets were compared to their abundances in different stages of planetary system formation. These studies found that the abundance ratios were consistent between the different source types and concluded that this indicated that the chemical compositions of the observed comets were likely inherited from the protosolar nebula \citep{Drozdovskaya2019, Lippi2024}. Isotopic fractionation ratios in comets can be indicative of the formation temperature of the molecule \citep{Rodgers2002, Mumma2011, Oberg2023} and have been employed in several investigations in the Solar System, such as the D/H fractionation ratio of \ce{H2O} \citep{Cleeves2014, Cordiner2025} or the \ce{^{13}C}/\ce{^{12}C} fractionation in polycyclic hydrocarbons (PAHs) in asteroids \citep{Zeichner2023}. Both studies indicated that in order to reproduce the isotopic ratios, the studied compounds may have been inherited from an earlier evolutionary stage, such as the parent molecular cloud or protoplanetary disk. However, since neither the molecular abundance nor isotope fractionation ratios are fixed, but rather can change as a result of certain environmental factors, the ratios alone are not sufficient evidence to make solid conclusions regarding inheritance. Additional information, such as whether the new conditions are able to produce the molecules in the quantities observed, is needed in order to draw these conclusions. Building up more complex and robust chemical inventories of comets, as well as of various stages of planetary system formation, will undoubtedly prove invaluable. 

The best methods for determining the composition of a comet nucleus are \textit{in situ} measurements, especially those that observe spectra at high spatial resolutions or analyze physical samples of the nucleus \citep{Mumma2011}{, such as the Rosetta mission to comet 67P/Churyumov-Gerasimenko \citep{Taylor2017, Filacchione2019, Rubin2019, Keller2020}}. As \textit{in situ} methods are very expensive{ and can only be carried out in a small sample of individual objects}, Earth-based telescopes are often used as an alternative. However, due to the small angular size of comet nuclei, Earth-based radio, millimeter, and infrared observations are necessarily sampling the gas-phase species in the coma as opposed to the rocky and icy nucleus \citep{Rodgers2004}. The gas-phase species in the coma can either be released through direct sublimation from the comet nucleus, which are referred to as parent species, or formed through photolysis of the parent species, referred to as daughter species. Without direct observations of the nucleus, parent species are the next best estimation of the nuclear composition \citep{Meech2004, Rodgers2004}.

On 2022 March 02, comet C/2022 E3 (ZTF) (hereafter referred to as E3) was detected by the Zwicky Transient Facility (ZTF) \citep{Bolin2022}. E3 is a long-period comet that likely originated in the Oort Cloud. E3's perihelion was on 2023 January 12, at which point it was 1.112 au from the sun, and it made its closest approach to Earth on 2023 February 01 at 0.284 au\footnote{\url{https://ssd.jpl.nasa.gov/tools/sbdb_lookup.html}}\textsuperscript{,}\footnote{\url{https://ssd.jpl.nasa.gov/horizons/app.html}}. Optical observations taken shortly after the perihelion determined {that }E3 has a rotation period of $\sim$8.5 h \citep{Knight2023, Manzini2023}. Follow-up observations of E3 with the ZTF determined that the abundances relative to CN of both \ce{C2} and \ce{C3} were comparable to those of other long-period solar system comets and that its dust production rate was typical of comets at similar heliocentric distances \citep{Bolin2024a}. From the same ZTF data, \cite{Liu2024} investigated the dust activity of E3 and estimated a nucleus size range of 0.81 $\pm$ 0.07 to 2.79 $\pm$ 0.01 km. Water production rates of E3 were measured with Odin shortly after its perihelion (UT 19{\textendash}20 Jan. 2023) by \cite{Biver2024a}, which also performed single-dish spectroscopic observations with the IRAM-30m, taken shortly after E3's closest approach to Earth (UT 03{\textendash}07 Feb. 2023). Comparisons between these results and those found in this paper {are discussed below}.

In this work, E3 was characterized with contemporaneous observations from both JWST and the Atacama Large Millimeter/submillimeter Array (ALMA). Multi-wavelength studies that sample different regions of the electromagnetic spectrum are able to search for molecules that may only be visible to one of the instruments. Contemporaneous observations allow for a more direct comparison of molecules that are visible to both instruments. In the case of cometary observations, this synergy proves to be invaluable for several reasons. For instance, ALMA requires optimal observing conditions and strong \ce{H2O} emission to observe water in the coma and is unable to sample the symmetric hydrocarbons (e.g., \ce{CH4}, \ce{C2H6}), whereas these are easily detected using JWST. Combined with the contemporaneity of the multi-wavelength observations, this means that the abundances of molecules detected by ALMA can be calculated with respect to a more accurate abundance of \ce{H2O}. Together, these techniques allow for the discovery of more robust chemical inventories for each comet observed. Section \ref{observations} describes the observations conducted, Section \ref{data-analysis} outlines the steps taken to analyze the data, and the results are presented in Section \ref{results} and discussed in Section \ref{discussion}.

\section{OBSERVATIONS} \label{observations}

Observations of E3 were made with ALMA (PI: M. Cordiner, ALMA\#2022.1.00997.T) and JWST (PI: S. Milam, PID: 1253) in 2023 February{\textendash}April, when the comet was between 1.3{\textendash}1.8 au from the Sun (0.9{\textendash}2.26 au from Earth). The Solar phase angle (Sun-Target-Observer) ranges from 48$^\circ$ to 45$^\circ$ over the course of the observations. Not all of the observations in the projects are discussed in this paper. Table \ref{table_obs} summarizes the observations in March that are to be discussed.

\begin{table*}
    \caption{\small Summary of the observations made of E3 with both JWST and ALMA in March of 2023}
    \label{table_obs}
    \centering
    \begin{threeparttable}
        \begin{tblr}{width=\textwidth,
        row{1-3}={c,m}, cell{3-13}{1-2}={c}, cell{2-3}{2}={l}, cell{3}{9}={r}}
            \toprule
            Telescope & {Observation Date \\ \vspace{2pt} {[UT]}} & {$\lambda$/$\nu$ \\ \vspace{2pt} {[µm/GHz]\tnote{b}}} & {Res.\tnote{c} \\ \vspace{2pt} {[nm/kHz]\tnote{b}}} & {Int.\tnote{d} \\ \vspace{2pt} {[min]}} & {$\Delta$\tnote{a} \\ \vspace{2pt} {[au]}} & {r\textsubscript{h}\tnote{a} \\ \vspace{2pt} {[au]}} & {$\theta$\textsubscript{B}\tnote{e} \\ \vspace{2pt} {[$^{\prime\prime}$]}} & {PA\tnote{f} \\ \vspace{2pt} {[$^\circ$]}}\\
            \toprule
            {JWST \\ \vspace{2pt} NIRSpec} & {2023-03-01 04:33{\textendash}05:07 \\ \vspace{2pt} 2023-03-01 07:37{\textendash}08:00} & {2.87{\textendash}5.27 \\ 1.66{\textendash}3.17} & {0.665 \\ \vspace{2pt} 0.396} & {46.7 \\ \vspace{2pt} 23.3} & {0.9398 \\ \vspace{2pt} 0.9436} & {1.3470 \\ \vspace{2pt} 1.3480} & 0.1$\times$0.1 & {\textendash}\\
            \midrule
            {ALMA \\ \vspace{2pt} Band 6} & {2023-03-02 00:19{\textendash}01:12 \\ \vspace{2pt} 2023-03-02 01:23{\textendash}02:16 \\ \vspace{2pt} 2023-03-02 23:40{\textendash}00:32 \\ \vspace{2pt} 2023-03-03 20:30{\textendash}21:21\tnote{g} \\ \vspace{2pt} 2023-03-03 21:33{\textendash}22:25\tnote{g} \\ \vspace{2pt} 2023-03-04 00:58{\textendash}01:50 \\ \vspace{2pt} 2023-03-04 22:01{\textendash}22:54\tnote{g}} & 241.63{\textendash}243.50 & 488.281 & 44.6 & {0.9686 \\ \vspace{2pt} 0.9700 \\ \vspace{2pt} 0.9979 \\ \vspace{2pt} 1.0240 \\ \vspace{2pt} 1.0254 \\ \vspace{2pt} 1.0296 \\ \vspace{2pt} 1.0560} & {1.3543 \\ \vspace{2pt} 1.3547 \\ \vspace{2pt} 1.3629 \\ \vspace{2pt} 1.3707 \\ \vspace{2pt} 1.3711 \\ \vspace{2pt} 1.3724 \\ \vspace{2pt} 1.3804} & {0.97$\times$0.68 \\ \vspace{2pt} 1.16$\times$0.63 \\ \vspace{2pt} 0.90$\times$0.70 \\ \vspace{2pt} 0.98$\times$0.76 \\ \vspace{2pt} 0.79$\times$0.76 \\ \vspace{2pt} 1.09$\times$0.66 \\ \vspace{2pt} 0.78$\times$0.73} & {-52.3 \\ \vspace{2pt} -59.4 \\ \vspace{2pt} -49.1 \\ \vspace{2pt} 80.0 \\ \vspace{2pt} -19.4 \\ \vspace{2pt} -59.0 \\ \vspace{2pt} -31.0}\\
            \bottomrule            
        \end{tblr}
        \begin{tablenotes}
            \item [a] \footnotesize $\Delta$ (distance between the comet and the observer) and r\textsubscript{h} (Heliocentric distance) values from NASA's JPL-Horizons Ephemeris System\footnote[3]{\url{https://ssd.jpl.nasa.gov/horizons/app.html}}
            \item [b] \footnotesize the spectral axes of JWST observations are shown in wavelength ($\lambda$; nm, µm) and of ALMA observations are shown in frequency ($\nu$; kHz, GHz)
            \item [c] \footnotesize spectral resolution
            \item [d] \footnotesize integration time
            \item [e] \footnotesize pixel size for JWST and resolving beam size for ALMA generated with natural weighting and convolved with a larger beam
            \item [f] \footnotesize beam position angle from CASA
            \item [g] \footnotesize observation was performed during the day {(between UTC 10:37{\textendash}23:24)}
        \end{tablenotes}
    \end{threeparttable}
\end{table*}

Near-infrared observations were made on UT 2023 March 01 with the Near-Infrared Spectrograph (NIRSpec) Integral Field Unit (IFU) on JWST. A four-point dither pattern was utilized for background subtraction. These observations used three disperser/filter pairings (G140H/F100LP, G235H/F170LP, and G395H/F290LP) to sample nearly the entire NIRSpec wavelength range from 1.0{\textendash}5.3 µm. The full width at half maximum (FWHM) of the point-spread functions (PSF) of the NIRSpec IFU increase by nearly an order of magnitude over the full bandwidth, though the pixel size is set by the NIRSpec instrument to be $0.1^{\prime\prime}\times 0.1^{\prime\prime}$. The disperser/filter pairs each have a resolving power of $\sim$2700, covering wavelength ranges 0.97{\textendash}1.89 µm, 1.66{\textendash}3.17 µm, and 2.87{\textendash}5.27 µm, respectively. These wavelength ranges contain several very strong \ce{H2O} vibrational emission bands \citep{Ellis1931, Bonner1934, Biver2024b}, as well as emission from \ce{CH3OH} \citep{Shimanouchi1972, Biver2024b}. Mid-infrared observations were also made of E3 on UT 2023 February 28 with the Mid-Infrared Instrument (MIRI) medium resolution spectrometer (MRS) on JWST, which covers a wavelength range of 4.9{\textendash}17.7 µm. The G140H/F100LP observation {(0.97\textendash 1.89 µm)} and the MIRI observations are outside the scope of this paper and are discussed in S. Milam et al. (in prep.).

Millimeter-wavelength observations of E3 were conducted with ALMA on UT 2023 March 02-04, as well as 2023 April 18, though at this point the comet was too distant to detect any \ce{CH3OH} lines. The March 2 observations utilized forty-three 12m antennas, while the rest utilized forty-four, in the moderately compact configuration C{\textendash}4. Though the observations each had several spectral windows that sample ALMA Band 6, this study specifically used a spectral window that covers the frequency range 241.63{\textendash}243.50 GHz. This spectral window had a channel width of 488.281 kHz and a median synthesized beam size of $\theta_B=0.7^{\prime\prime}\times 0.5^{\prime\prime}$. This frequency range contains several detected transitions from the \ce{CH3OH} $J=5_{K^\prime_a,K^\prime_c}-4_{K^{\prime\prime}_a,K^{\prime\prime}_c}$ transition ladder \citep{Endres2016}, which are shown in Table \ref{table_lines}.

\begin{table}[]
    \caption{Parameters\textsuperscript{a}\tnote{a} of the detected $J=5_{K^\prime_a,K^\prime_c}-4_{K^{\prime\prime}_a,K^{\prime\prime}_c}$ \ce{CH3OH} rotational transitions observed with ALMA}
    \label{table_lines}
    \begin{threeparttable}
        \begin{tblr}{width=\columnwidth,
        colspec={X[-1,c]X[-1,c]X[-1,c]X[-1,c]},
        row{1}={m}, cell{2-18}{1}={mode=math}}
            \toprule
            Transition & {Frequency\tnote{b} \\ \vspace{2pt} {[GHz]}} & {E\textsubscript{u} \\ \vspace{2pt} {[K]}} & $\log_{10}A_{ij}$ \\
            \toprule
            5_{0,5}-4_{0,4} \:E & 241.700159 & 47.9 & -4.21933 \\
            5_{-1,5}-4_{-1,4} \:E & 241.767234 & 40.4 & -4.23614 \\
            5_{0,5}-4_{0,4} \:A^+ & 241.791352 & 34.8 & -4.21856 \\
            5_{-4,2}-4_{-4,1} \:A^- & 241.806524 & 115.2 & -4.66153 \\
            5_{4},1-4_{4,0} \:A^+ & 241.806525 & 115.2 & -4.66153 \\
            5_{-4,2}-4_{-4,1} \:E & 241.813255 & 122.7 & -4.66207 \\
            5_{4,1}-4_{4,0} \:E & 241.829629 & 130.8 & -4.65962 \\
            5_{3,3}-4_{3,2} \:A^+ & 241.832718 & 84.6 & -4.4127 \\
            5_{-3,2}-4_{-3,1} \:A^- & 241.833106 & 84.6 & -4.41269 \\
            5_{-2,4}-4_{-2,3} \:A^- & 241.842284 & 72.5 & -4.29118 \\
            5_{3,2}-4_{3,1} \:E & 241.843604 & 82.5 & -4.4114 \\
            5_{-3,3}-4_{-3,2} \:E & 241.852299 & 97.5 & -4.40947 \\
            5_{1,4}-4_{1,3} \:E & 241.879025 & 55.9 & -4.22473 \\
            5_{2,3}-4_{2,2} \:A^+ & 241.887674 & 72.5 & -4.2089 \\
            5_{-2,4}-4_{-2,3} \:E & 241.904147 & 60.7 & -4.29306 \\
            5_{2,3}-4_{2,2} \:E & 241.904643 & 57.1 & -4.29845 \\
            \midrule
       \end{tblr}
       \begin{tablenotes}
           \item [a] \footnotesize spectral parameters are from the Cologne Database for Molecular Spectroscopy \citep{Endres2016}
           \item [b] \footnotesize error for all frequencies is 4 kHz
       \end{tablenotes}
   \end{threeparttable}
\end{table}

Depending on the observational set-up and ALMA imaging parameters, ALMA and JWST observations of the same object can have relatively comparable spatial resolution. For E3, the pixel sizes are very similar{\textemdash}68 km (0.1$^{\prime\prime}$) for the JWST observations, set by the NIRSpec detectors, and 59 km (0.08$^{\prime\prime}$) for the ALMA observations, such that the synthesized beam prior to the JvM correction covers $\sim$5 pixels (the JvM effect and subsequent correction are discussed in Section \ref{data-analysis}). The resolving beams for each ALMA observation following the JvM correction are presented in Table \ref{table_obs}, with an average beam size of {$\theta_B=0.95\times0.70^{\prime\prime}$.} Due to the higher frequency of the JWST measurements, the wavelength dependence of the FWHM of the PSF is significant, spanning nearly an order of magnitude over the full spectral range of NIRSpec. For the two filters discussed in this work, the roughly circular PSFs have FWHM values that range from 0.055$^{\prime\prime}$ at 1.66µm to 0.17$^{\prime\prime}$ at 5.27µm. While the JWST PSF FWHM are all smaller than the ALMA restoring beams, the pixel size is set by the NIRSpec instrument to be similar to the PSF FWHM near the center of the entire NIRSpec spectral band. This pixel size undersamples the PSF even at the largest wavelengths, though dithering can improve the PSF sampling \citep{Jakobsen2022}. The ALMA observations have an average maximum recoverable scale of $\sim$8,800 km ($\sim$12$^{\prime\prime}$) and a field-of-view with a radius of $\sim$13,000 km ($\sim$18$^{\prime\prime}$). This is larger than the JWST observations, which span up to $\sim$2075 km ($\sim$3$^{\prime\prime}$). The achievable spectral resolving power of ALMA (R$=\sim$500,000) is orders of magnitude higher than that of JWST NIRSpec (R$=\sim$2700), and allows for the characterization of the outgassing kinetics in the coma through Doppler shifting. As an interferometer, ALMA is comprised of approximately 2500 pairs of telescopes which allow the array to sample a range of spatial scales. These observations can be combined with radiative transfer modeling to determine detailed three-dimensional spatial distributions of volatiles, which can be compared to the profiles of direct nucleus sublimation and coma production to infer the source of the volatile \citep{Cordiner2023}.

\section{DATA ANALYSIS} \label{data-analysis}

\begin{table*}[]
    \centering
    \caption{Descriptions of highlighted spectral regions of the JWST spectra in Figure \ref{fig_Jspec}}
    \label{table_jwstspec}
    \begin{threeparttable}
        \begin{tblr}{
        cell{1-6}{1-3}={valign=m}, cell{1-5}{4}={valign=m},
        cell{6}{4}={valign=f},
        cell{1}{5-6}={valign=m},
        cell{2-6}{5-6}={valign=f},
        colspec={X[-1,c] X[-1,c] X[-1,c] X[-1,c] X[-1,c] X[c]}}
            \toprule
            Label & {Wavelength \\ Range [µm]} & {Disperser-Filter \\ Pairing} & Molecule & Vibrational Mode & Transition Description \\
            \toprule
            {[a]} & 2.56{\textendash}2.80 & G235H-F170LP & \ce{H2O} & {$\nu_3$ 1$\xrightarrow{\hspace*{0.2cm}}$0 \\ $\nu_1$ 1$\xrightarrow{\hspace*{0.2cm}}$0} & {fundamental asymmetric stretch\tnote{a}\textsuperscript{,}\tnote{g} \\ fundamental symmetric stretch\tnote{a}\textsuperscript{,}\tnote{g}} \\
            \midrule
            {[b]} & 2.85{\textendash}3.00 & G235H-F170LP & \ce{H2O} & {$\nu_1$ 1$\xrightarrow{\hspace*{0.2cm}}$0 \\ $\nu_1 + \nu_3\xrightarrow{\hspace*{0.2cm}}\nu_1$ \\ $2\nu_1\xrightarrow{\hspace*{0.2cm}}\nu_3$} & {fundamental symmetric stretch\tnote{g} \\ hot band\tnote{g} \\ combination band (difference transition)\tnote{g}} \\
            \midrule
            {[c]} & 2.87{\textendash}3.00 & G395H-F290LP & \ce{H2O} & {$\nu_1$ 1$\xrightarrow{\hspace*{0.2cm}}$0 \\ $\nu_1 + \nu_3\xrightarrow{\hspace*{0.2cm}}\nu_1$ \\ $2\nu_1\xrightarrow{\hspace*{0.2cm}}\nu_3$} & {fundamental symmetric stretch\tnote{g} \\ hot band\tnote{g} \\ combination band (difference transition)\tnote{g}} \\
            \midrule
            {[d]} & 3.48{\textendash}3.56 & G395H-F290LP &  \ce{CH3OH} & $\nu_3$ 1$\xrightarrow{\hspace*{0.2cm}}$0 & fundamental \ce{CH3} symmetric stretch\tnote{a}\textsuperscript{,}\tnote{f} \\
            \midrule
            {[e]} & 4.50{\textendash}5.23 & G395H-F290LP & {\ce{H2O} \\ \ce{H2O} \\ \ce{CO} \\ \ce{CN} \\ \ce{OCS}} & {$\nu_3\xrightarrow{\hspace*{0.2cm}}\nu_2$ \\ $\nu_1\xrightarrow{\hspace*{0.2cm}}\nu_2$ \\ 1$\xrightarrow{\hspace*{0.2cm}}$0 \\ 1$\xrightarrow{\hspace*{0.2cm}}$0 \\ $\nu_1$ 1$\xrightarrow{\hspace*{0.2cm}}$0} & {combination band (difference transition)\tnote{b}\textsuperscript{,}\tnote{g} \\ combination band (difference transition)o\tnote{b}\textsuperscript{,}\tnote{g} \\ fundamental band\tnote{e} \\ fundamental band\tnote{d} \\ fundamental CO stretch\tnote{c}} \\
            \bottomrule
        \end{tblr}
        \begin{tablenotes}
           \item [a] \footnotesize \cite{Shimanouchi1972}
           \item [b] \footnotesize \cite{DelloRusso2000}
           \item [c] \footnotesize \cite{Tubergen2000}
           \item [d] \footnotesize \cite{Horka2004}
           \item [e] \footnotesize \cite{Villanueva2011}
           \item [f] \footnotesize \cite{Villanueva2012a}
           \item [g] \footnotesize \cite{Villanueva2012b}
       \end{tablenotes}
    \end{threeparttable}
\end{table*}

The JWST NIRSpec data were downloaded from MAST (\dataset[https://doi.org/10.17909/czbe-fk30]{https://doi.org/10.17909/czbe-fk30}) and were calibrated by the 1.16.0 release of the JWST Science Calibration Pipeline \citep{jwst} and the resulting FITS cubes were inspected using \texttt{CubeViz}, a sub-module of \texttt{jdaviz} \citep{jdaviz}.

The spectrum was extracted for each pixel and spectral regions of interest (discussed in detail in Table \ref{table_jwstspec}) were modeled to determine the best-fit rotational temperatures and production rates of \ce{H2O} and \ce{CH3OH} using NASA's Planetary Spectrum Generator (PSG) \citep{Villanueva2017}. 

The ALMA Band 6 data were flagged and calibrated using the ALMA Science Pipeline in the Common Astronomy Software Applications (CASA) package \citep{casa}. The calibrated measurement sets were then continuum subtracted and imaged with the CASA \texttt{tclean} task. The deconvolution and weighting parameters in \texttt{tclean} were set to H\"ogbom and natural, respectively, and the pixel size was set to $\sim$0.08$^{\prime\prime}$. In addition to the images that were CLEANed from each individual March observation, a combined image was generated by concatenating all of the March observation measurement sets into one image prior to cleaning.

The point-spread functions (PSFs) of these observations were significantly non-Gaussian, and residual scaling was performed following \cite{Czekala2021} to correct for the Jorsater \& van Moorsel (JvM) effect. When a PSF is significantly non-Gaussian, it will have "shelves" of emission in the outer regions that will not be accounted for by the CLEAN beam. This results in a CLEAN beam that has a significantly smaller volume than the PSF, resulting in CLEANed images with lower fluxes. Local Thermodynamic Equilibrium (LTE) modeling of spectra relies heavily on the relative intensities of lines, so it is critical that this effect be corrected for. The JvM correction is done by first calculating the degree of non-Gaussianity, $\epsilon$, from the volume ratio between the CLEAN and dirty beams. The residual image from \texttt{tclean} are then scaled by $\epsilon$ and added to the model image, which has been convolved with the CLEAN beam. For additional flux recovery, but lower spatial resolution, the model image can be convolved with a larger resolving beam. For these ALMA measurements, $\epsilon$ ranged from 0.51 to 0.58. In order to further improve the signal-to-noise of the images an additional convolution was performed, increasing the CLEAN beam by a factor of $\epsilon^{-1/2}$.

Spectra were then extracted from the images and fit using the LTE model from \cite{Cordiner2017}. This LTE model fits an extracted spectrum using the Python non-linear least squares fitting package \texttt{mpfit} \citep{mpfit}, varying the Doppler shift, coma expansion velocity, column density, and rotational temperature. The model assumes a Gaussian line shape and accounts for optical depth effects. This procedure was applied to the spectrum extracted from each pixel in a circular region with a 20-pixel radius around the continuum photocenter of the comet to fit the rotational temperature and column density of \ce{CH3OH}. As the modeled regions are at most 1270 km in radius, they are sampling the innermost area of the coma, where, due to the small heliocentric distance and high \ce{H2O} gas density, there are higher collision rates in the gas that couple the kinetic and rotational excitation temperatures and LTE conditions are assumed to be met. Primary beam correction was not employed as these circular apertures are close enough to the center of the image for the change in flux density to be negligible with respect to the average root mean square noise (RMS) in the image. 

\begin{sidewaysfigure*}
    \centering
    \includegraphics[width=\linewidth]{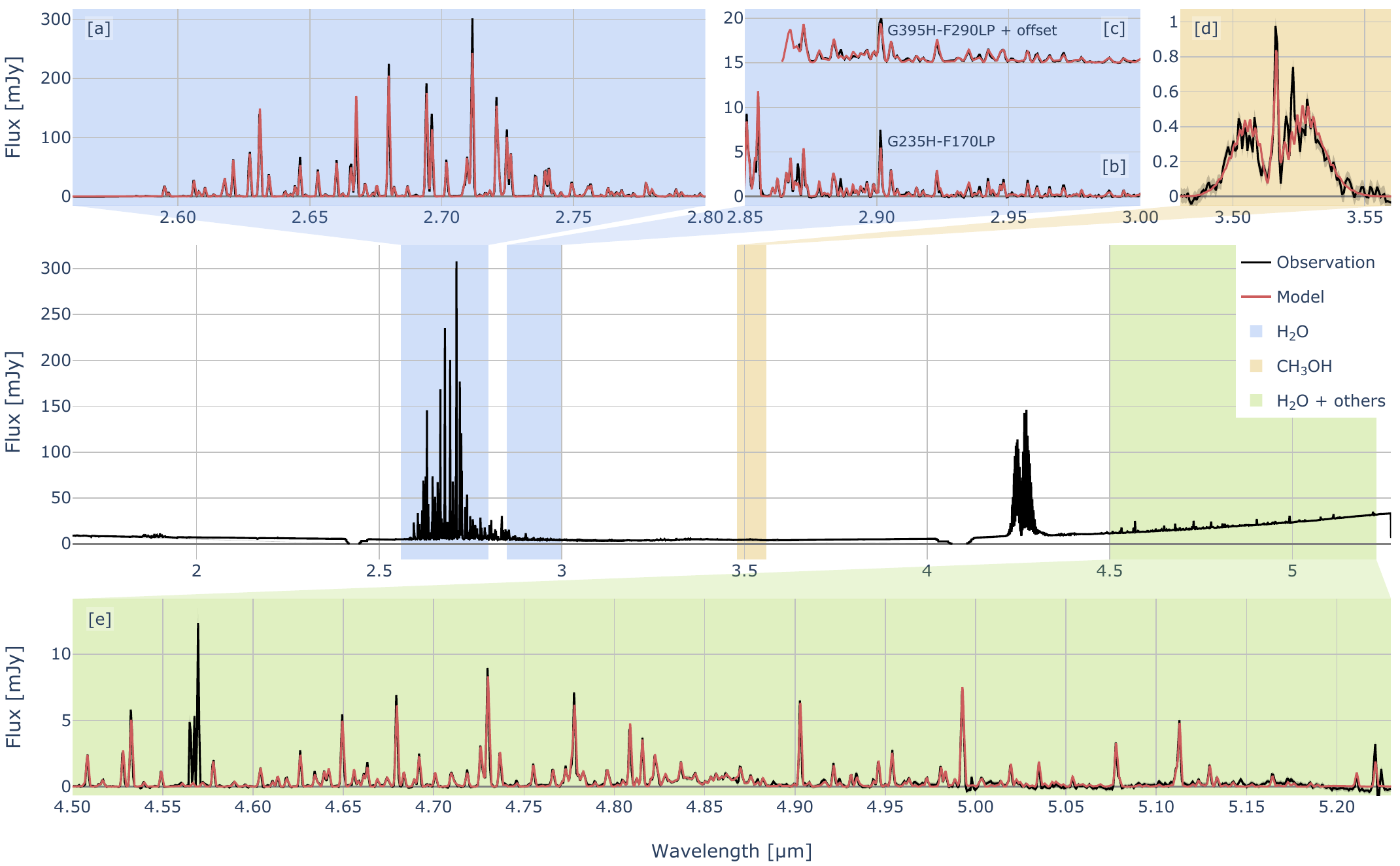}
    \caption{(center) JWST NIRSpec spectrum for each of the two filter-grating pairings: G235H-F170LP (1.66–3.17 µm) and G395H-F290LP (2.87–5.27 µm) extracted from a 0.47$^{\prime\prime}$-radius circular aperture centered on the continuum photocenter. Shaded regions show the wavelength regions used in PSG modeling. {Subplots above and below the center spectra show zoomed-in regions of the spectra that are baseline-subtracted along with the PSG model spectra overlaid. The blue subplots (a, b, c) are wavelength regions dominated by emission from \ce{H2O}, while the orange subplot (d) shows the wavelength region with emission from \ce{CH3OH}, and the green subplot (e) below the center spectra contains emission from \ce{H2O} as well as other molecules.} For further discussion of the wavelength regions, see Table \ref{table_jwstspec}.}
    \label{fig_Jspec}
\end{sidewaysfigure*}

Of the seven ALMA observations made in March, three were made during the day, at most starting {nearly 3} hours before sunset, as indicated in Table \ref{table_obs}. This resulted in increased variability of the amplitude calibrators due to a significant increase in the phase variation as a result of the decreased phase stability of the atmosphere from solar irradiation. To mitigate these effects, automated self-calibration was attempted, but the signal-to-noise ratio of the continuum images was too low to allow any self-calibration to be performed. As a result, the modeled values for rotational temperature and column density of these observations have larger uncertainties. 

\begin{table*}[]
    \centering
    \caption{Modeled physical parameters of E3}
    \label{table_physical}
    \begin{threeparttable}
        \begin{tblr}{
        colspec={cX[c]cccccc},
        row{1}={m},
        cell{2}{1-3}={r=3}{c},
        cell{5}{1-3}={r=2}{c}}
            \toprule
            {Obs. Date \\ \vspace{2.5pt} {[UT]}} & Telescope & {r\textsubscript{h}\tnote{a} \\ \vspace{2.5pt} {[au]}} & Mol. & {JWST \\ Spectral \\ Region} & {T\textsubscript{rot} \\ \vspace{2.5pt} {[K]}} & {N[Mol.] \\ \vspace{2.5pt} {[$\times$10\textsuperscript{17} m\textsuperscript{-2}]}} & {Q[Mol.] \\ \vspace{2.5pt} {[$\times$10\textsuperscript{26} molec.~s\textsuperscript{-1}]}} \\
            \toprule
            2023-03-01 04:33 & JWST\tnote{b} & 1.3470 & \ce{H2O} & [c] & 68.8 $\pm$ 1.7 & 440. $\pm$ 10.5 & 359 $\pm$ 8.60 \\
            &  &  & \ce{H2O} & [e] & 65.6 $\pm$ 1.3 & 442 $\pm$ 5.55 & 360. $\pm$ 4.53 \\
            &  &  &  \ce{CH3OH} & [d] & 47.1 $\pm$ 2.2 & 4.87 $\pm$ 0.37 & 3.97 $\pm$ 0.30 \\
            \hline[dotted]
            2023-03-01 07:37 & JWST\tnote{b} & 1.3480 & \ce{H2O} & [a] & 68.3 $\pm$ 0.7 & 302 $\pm$ 2.86 & 248 $\pm$ 2.35 \\
            &  &  & \ce{H2O} & [b] & 75.0 $\pm$ 1.8 & 326 $\pm$ 5.52 & 268 $\pm$ 4.52 \\
            \midrule
            2023-03-02 00:19 & ALMA & 1.3543 & \ce{CH3OH} & {\textendash} & 64.5 $\pm$ 8.2 & 9.36 $\pm$ 1.00 & 3.90 $\pm$ 0.42 \\
            2023-03-02 01:23 & ALMA & 1.3547 & \ce{CH3OH} & {\textendash} & 59.4 $\pm$ 0.4 & 9.05 $\pm$ 0.28 & 3.74 $\pm$ 0.12 \\
            2023-03-02 23:40 & ALMA & 1.3629 & \ce{CH3OH} & {\textendash} & 44.4 $\pm$ 4.0 & 7.85 $\pm$ 0.42 & 3.34 $\pm$ 0.18 \\
            2023-03-03 20:30 & ALMA & 1.3707 & \ce{CH3OH} & {\textendash} & 87.9 $\pm$ 7.3 & 10.46 $\pm$ 0.96 & 4.98 $\pm$ 0.46 \\
            2023-03-03 21:33 & ALMA & 1.3711 & \ce{CH3OH} & {\textendash} & 74.8 $\pm$ 12.2 & 15.88 $\pm$ 2.78 & 6.89 $\pm$ 1.21 \\
            2023-03-04 00:58 & ALMA & 1.3724 & \ce{CH3OH} & {\textendash} & 44.4 $\pm$ 1.5 & 7.24 $\pm$ 0.06 & 3.25 $\pm$ 0.03 \\
            2023-03-04 22:01 & ALMA & 1.3804 & \ce{CH3OH} & {\textendash} & 37.9 $\pm$ 1.6 & 9.25 $\pm$ 0.28 & 4.04 $\pm$ 0.12 \\
            Combined ALMA & ALMA & 1.3667 & \ce{CH3OH} & {\textendash} & 56.1 $\pm$ 0.8 & 9.70 $\pm$ 0.04 & 4.15 $\pm$ 0.02 \\
            \midrule
            ALMA Average\tnote{c} & ALMA & 1.3667 & \ce{CH3OH} & {\textendash} & 59.0 $\pm$ 17.1 & 9.87 $\pm$ 3.16 & 4.30 $\pm$ 1.38 \\
            \ce{H2O} Average & JWST\tnote{b} & 1.3475 & \ce{H2O} & [a-c,e] & 69.4 $\pm$ 2.9 & 378 $\pm$ 13.4 & 309 $\pm$ 11.0 \\
            \ce{CH3OH} Average\tnote{c}\textsuperscript{,}\tnote{d} & JWST\tnote{b} + ALMA & 1.3609 & \ce{CH3OH} & [d] & 57.5 $\pm$ 17.2 & 9.24 $\pm$ 3.18 & 4.26 $\pm$ 1.41 \\
            \bottomrule
        \end{tblr}
        \begin{tablenotes}
            \item[a] \footnotesize heliocentric distances for combined images and averages are averaged over the entire time span
           \item [b] \footnotesize JWST values are modeled from spectra extracted from a circular aperture centered on the nucleus with a radius of 0.47\textsuperscript{$\prime\prime$}
           \item [c] \footnotesize averages do not include the combined ALMA image values, only values derived from individual observations
           \item [d] \footnotesize overall average determined from values modeled only from \ce{CH3OH} 
        \end{tablenotes}
    \end{threeparttable}
\end{table*}

\section{RESULTS} \label{results}

The JWST NIRSpec spectra for the two disperser-filter pairings are shown in Figure \ref{fig_Jspec}, which were extracted from an aperture around the continuum photocenter with the same area as the combined ALMA beam size, calculated to be a circular region with a 0.47$^{\prime\prime}$ radius. To model the infrared spectra {with the PSG \citep{Villanueva2017}}, subsets of the spectra were taken from wavelength regions with strong vibrational emission from specific molecules. {These model spectra are then plotted along with the baseline-subtracted spectra.} The shaded regions in Figure \ref{fig_Jspec} depict these spectral subsets and the molecular vibrational emission of each subset is described in Table \ref{table_jwstspec}. {Any emission bands outside of the spectral regions outlined in Table \ref{table_jwstspec} are outside of the scope of this paper. The unmodeled feature in region [e] at $\sim$4.57 µm appears to be an instrumental artefact.}

\begin{figure*}
    \centering
    \includegraphics[width=\linewidth]{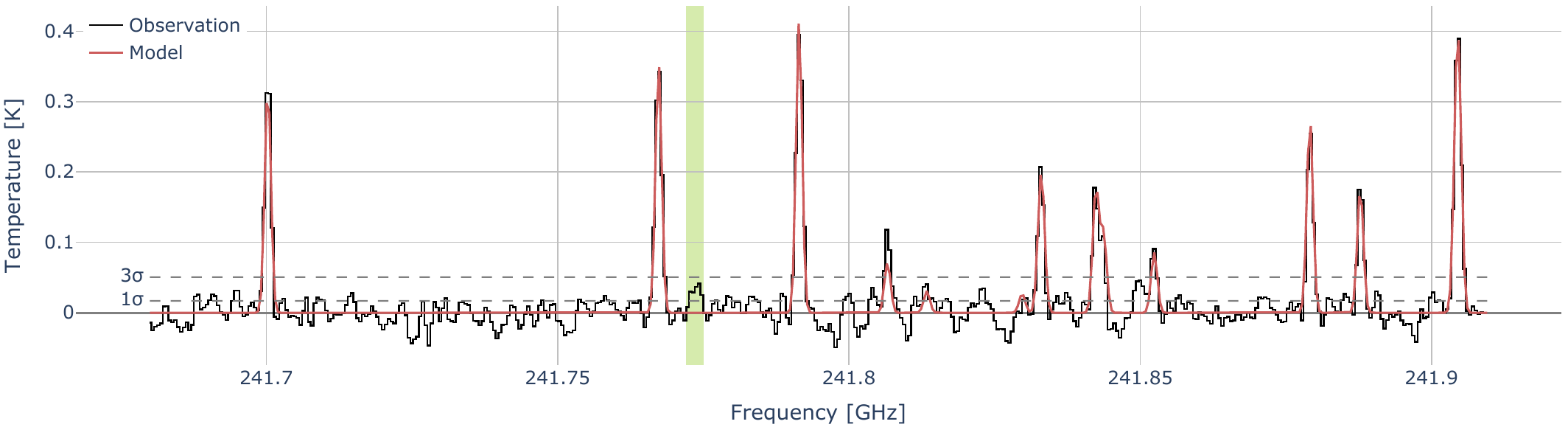}
    \caption{ALMA spectrum of \ce{CH3OH} extracted from the center of the methanol emission peak. The best-fit LTE model is overlaid in red, providing a modeled rotational temperature ($\mathrm{T_{rot}}$) of 56.11 ± 0.77 K and a methanol column density (N[\ce{CH3OH}]) of (9.70 ± 0.04)$\times$10\textsuperscript{17} m\textsuperscript{-2}. Dashed lines depict the thresholds for 1$\sigma$ (0.0169 K) and 3$\sigma$ (0.0507 K).{The potential \ce{HNCO} line is highlighted in green.}}
    \label{fig_Aspec}
\end{figure*}

{The main sources of continuum emission in comets are dust in the coma and, to a lesser extent, the nucleus. At shorter wavelengths (less than 3 µm), the continuum is dominated by reflected emission, while at wavelengths longer than 4 µm, thermal emission dominates. The PSG models the continuum based on the size and emissivity of the nucleus as well as the dust production rate. The nucleus size of E3 is not stringently constrained in the literature, and modeling the dust production rate is beyond the scope of this work. Instead, the continuum was fit with a polynomial baseline, which was used to generate the continuum-subtracted JWST spectra.} For each of the wavelength regions, a best-fit rotational temperature was determined, fitting for the physical parameters of the gas in the cometary coma. For wavelength regions containing \ce{H2O} emission (Figure 1 regions [a-c,e]), the total gas production rate was determined and for the wavelength region that contained emission from \ce{CH3OH} (Figure 1 region [d]), the molecular abundance relative to \ce{H2O} was determined. Given the total gas production rate and the relative molecular abundance, the production rate of \ce{CH3OH} was determined. These molecular production rates were then converted to column densities using a standard Haser model \citep{Haser1957} where the expansion velocity was assumed to be $0.8\times r_h$ km/s and the photoionization rates were determined assuming an active Sun from PHoto Ionization/Dissociation RATES\footnote[4]{https://phidrates.space.swri.edu/}. This modeling was carried out for every pixel, as well as for circular apertures with radii of 0.47$^{\prime\prime}$, a region with the same area as the combined ALMA beam. Similarly, for the ALMA observations, the \ce{CH3OH} $J=5_{K^\prime_a,K^\prime_c}-4_{K^{\prime\prime}_a,K^{\prime\prime}_c}$ spectrum extracted from each pixel of the ALMA Band 6 observations was modeled, both for each individual observation, as well as for a composite image generated in CASA from all of the March observations combined. This combined image was generated by concatenating all of the measurement sets from the March observations prior to CLEANing. Each extracted spectrum from the ALMA observations was fit for the same parameters for \ce{CH3OH}: rotational temperature and column density.{ A spectrum extracted from the combined ALMA image at the center of the \ce{CH3OH} emission is shown in Figure \ref{fig_Aspec}.} Table \ref{table_physical} contains the modeled physical parameters from these circular JWST apertures, as well as the modeled parameters from the \ce{CH3OH} emission peak in the ALMA observations.

\begin{figure*}
    \centering
    \includegraphics[width=\linewidth]{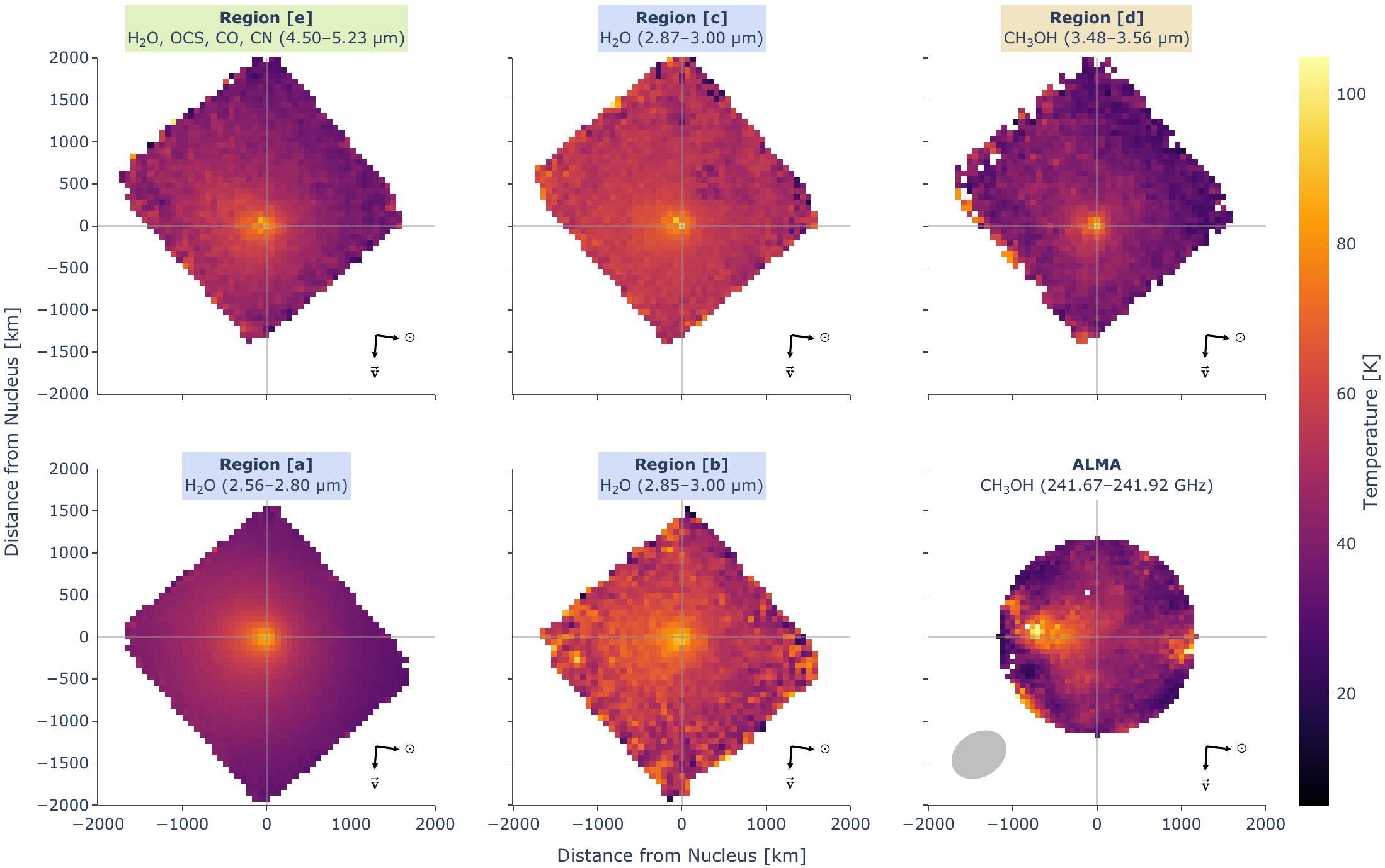}
    \caption{Modeled excitation temperature maps for the wavelength and frequency regions discussed in Figures \ref{fig_Jspec} and \ref{fig_Aspec}. {The rows split the JWST maps by observation: (top) G235H-F170LP and (bottom) G395H-F290LP, while the columns group the maps by molecule: (left) predominantly \ce{H2O}, (center) \ce{H2O}, specifically the spectral region covered by both observations, and (right) \ce{CH3OH}.} The maps are masked to only show values at 3$\sigma$ or greater confidence. Maps modeled from the JWST spectra are labeled with a letter corresponding to the region in Figure \ref{fig_Jspec}. The bottom-right map was modeled from the ALMA rotational spectrum (Figure \ref{fig_Aspec}). Arrows in the bottom-right corner indicate the direction of the Sun ($\odot$) and comet velocity ($\vec{\mathrm{v}}$). The ALMA synthesized beam is shown in the bottom-left corner.}
    \label{fig_JAtemps}
\end{figure*}

The resulting temperature maps are shown in Figure \ref{fig_JAtemps}. All of the temperature maps modeled from JWST spectral regions display a centrally peaked distribution and three of the \ce{H2O}-containing regions ([a,b,e]) have largely similar peak temperatures of $\sim$87 K ([a]: 87.2 $\pm$ 1.0 K; [b]: 88.2 $\pm$ 1.8 K; and [e]: 87.0 $\pm$ 5.7 K). Spectral region [c] has a higher peak modeled temperature of 97.0 $\pm$ 6.8 K that is closer to the peak temperatures of \ce{CH3OH} spectral region ([d]), 96.1 $\pm$ 5.5 K, and the combined March ALMA image, 103.5 $\pm$ 1.2 K. {This discrepancy is likely due to the relative sizes of the ALMA and JWST beams with respect to the pixel sizes. While the JWST beam and pixel sizes are relatively similar, the ALMA beam size is larger than a pixel, which causes emission from objects smaller than the beam to be smeared (see discussion in Section \ref{observations}). To account for this in comparing modeled results from both observational facilities, JWST values were determined by averaging modeled values over a circular aperture with the same area as the combined ALMA image beam. The} aperture temperatures for the \ce{H2O}-containing regions are all in general agreement, with an average of 69.4 $\pm$ 2.9 K, while the JWST \ce{CH3OH} spectral region has a lower aperture temperature of 47.1 $\pm$ 2.2 K. Unlike the JWST temperature maps, the temperature map for the combined March ALMA image doesn't peak at the continuum peak, but rather in the anti-sunward direction from the comet nucleus at a radial offset of 712 km, although the temperature at the \ce{CH3OH} emission peak, 56.1 $\pm$ 0.8 K, is similar to the JWST \ce{CH3OH} aperture value. 

\begin{figure*}
    \centering
    \includegraphics[width=0.695\linewidth]{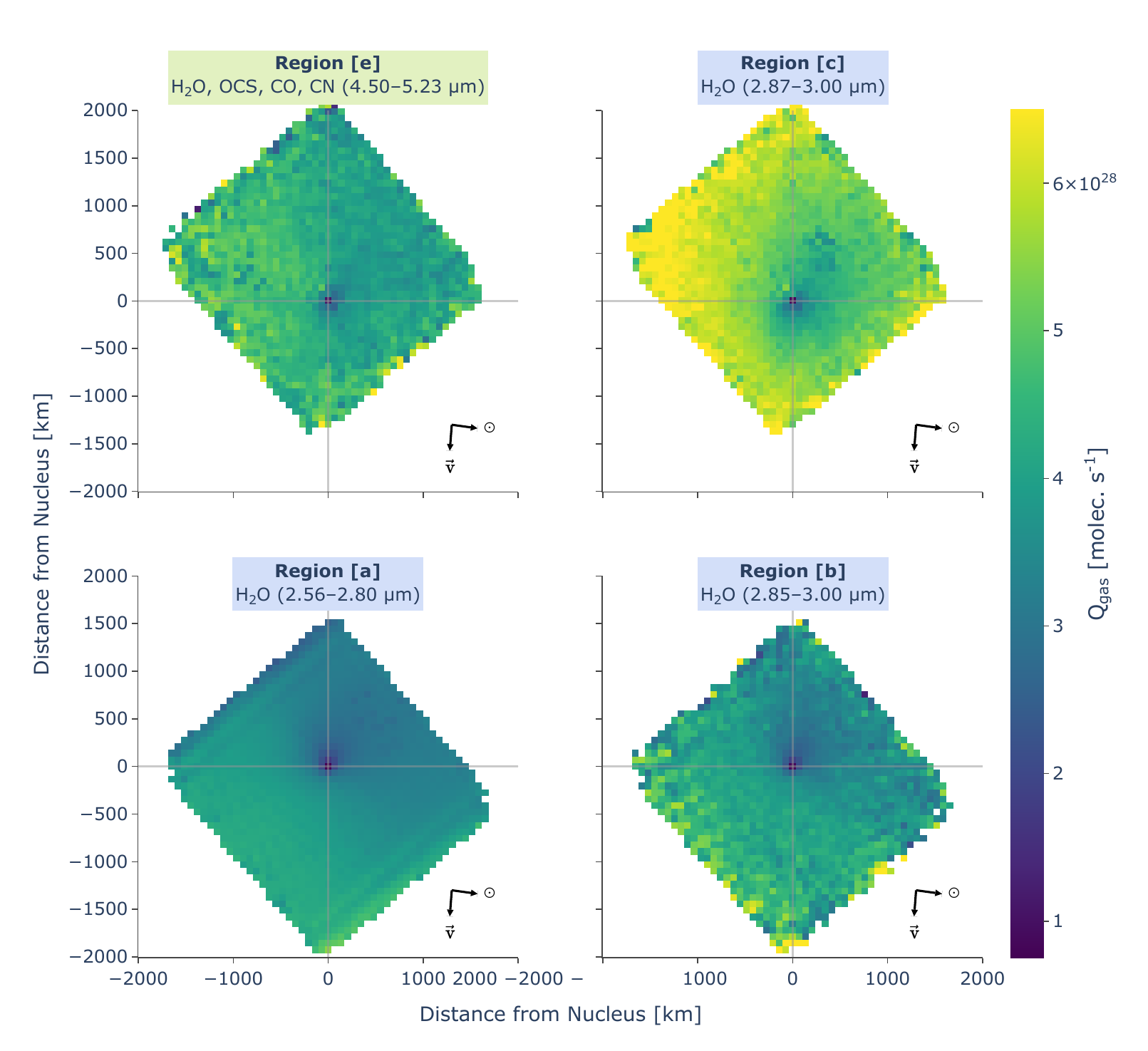}
    \caption{Maps depicting modeled gas production rates for the wavelength regions in Figure \ref{fig_Jspec} that contain \ce{H2O} ro-vibrational emission. {The rows split the maps by observation: (top) G235H-F170LP and (bottom) G395H-F290LP, while the columns group the maps by molecule: (left) predominantly \ce{H2O} and (right) \ce{H2O}, specifically the spectral region covered by both observations.} The maps are masked to only show values at 3$\sigma$ or greater confidence. Maps are labeled with a letter corresponding to the wavelength region. The top two maps are from the first observation and the bottom two are from the second observation, taken 3 hours later. Arrows in the bottom-right corner indicate the direction of the Sun ($\odot$) and comet velocity ($\vec{\mathrm{v}}$).}
    \label{fig_Jact}
\end{figure*}

\begin{figure*}
    \centering
    \includegraphics[width=0.71\linewidth]{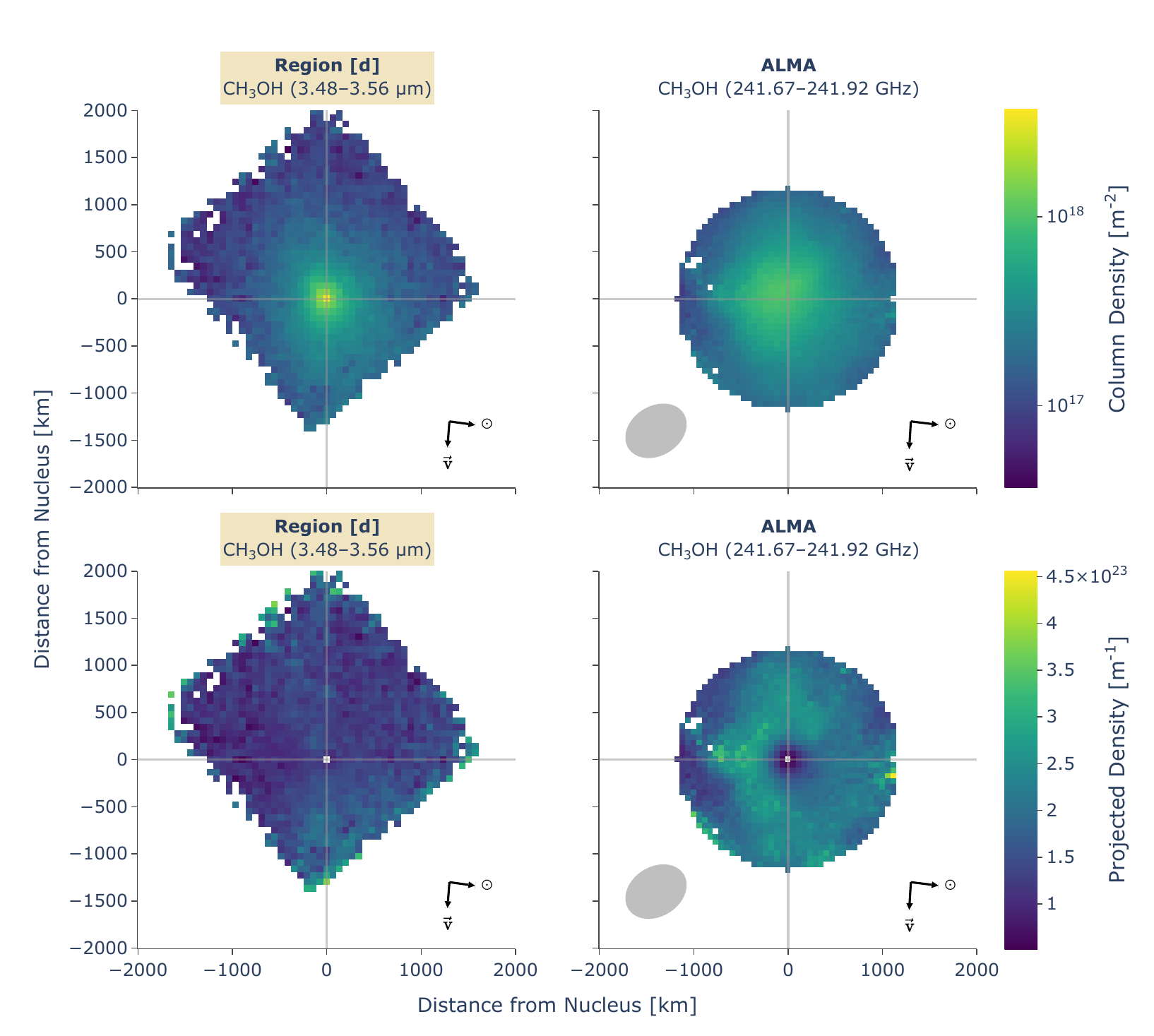}
    \caption{\ce{CH3OH} (top) column density maps and (bottom) projected density maps modeled from (left) the \ce{CH3OH} ro-vibrational emission peak in JWST region [d] (Figure \ref{fig_Jspec}) and (right) the rotational emission ladder from the ALMA observations (Figure \ref{fig_Aspec}). The maps are masked to only show values at 3$\sigma$ or greater confidence. Arrows in the bottom-right corner indicate the direction of the Sun ($\odot$) and comet velocity ($\vec{\mathrm{v}}$). The ALMA synthesized beam is shown in the bottom-left corner.}
    \label{fig_Jch3oh}
\end{figure*}

The modeled maps for the local gas production rates are shown in Figure \ref{fig_Jact}. Due to the opacity of \ce{H2O} near the comet nucleus and PSF effects, pixels near the center underestimate the actual gas production rate. Using the aperture values, the global gas production rate evolves from an average of {(360. $\pm$ 10)$\times$10\textsuperscript{26}} molec.~s\textsuperscript{-1} during the first observation to {(258 $\pm$ 5)$\times$10\textsuperscript{26}} molec.~s\textsuperscript{-1} during the second observation, with good agreement between different \ce{H2O} wavelength regions in the same observation. The \ce{CH3OH} column density maps are shown in Figure \ref{fig_Jch3oh} for both the JWST and ALMA observations. Both column density maps show a centrally peaked distribution, {although the column density peak derived from JWST spectral region [d], (36.9 $\pm$ 2.14)$\times$10\textsuperscript{17} m\textsuperscript{-2}, is approximately four times the value derived from the ALMA combined March cube, (9.70 $\pm$ 0.04)$\times$10\textsuperscript{17} m\textsuperscript{-2}. Similarly to the modeled temperature maps, the aperture \ce{CH3OH} column density from spectral region [d], (4.87 $\pm$ 0.37)$\times$10\textsuperscript{17} m\textsuperscript{-2}, is closer to the combined ALMA \ce{CH3OH} column density.} The JWST \ce{CH3OH} aperture production rate is (3.97 $\pm$ 0.30)$\times$10\textsuperscript{26} molec.~s\textsuperscript{-1}, which is in agreement with the production rate determined from the ALMA combined image, (4.15 $\pm$ 1.38)$\times$10\textsuperscript{26} molec.~s\textsuperscript{-1}. To calculate the \ce{CH3OH} abundance with respect to (wrt) \ce{H2O}, the JWST value will employ the \ce{H2O} production rate from spectral region [e] as the total gas production rate, as both spectral regions are from the same observation, while the combined ALMA value will employ the average \ce{H2O} production rate as the total gas production rate, as the \ce{H2O} production rates are variable on hourly timescales. These production rates correspond to relative abundances of \ce{CH3OH} wrt \ce{H2O} of 1.10 $\pm$ 0.08\% for JWST spectral region [d] and 1.34 $\pm$ 0.04\% for the combined ALMA image. The bottom panels in Figure \ref{fig_Jch3oh} show projected density, in which the column densities were {divided by 1/$\rho$, where $\rho$ is} the projected distance from the comet nucleus{,} to enhance any non-isotropic expansion of the coma. The projected density map for the JWST \ce{CH3OH} spectral region shows two faint jets, one to the south and the other slightly west of north, which are not in the projected density map for the ALMA combined image.

\begin{figure}
    \centering
    \includegraphics[width=\columnwidth]{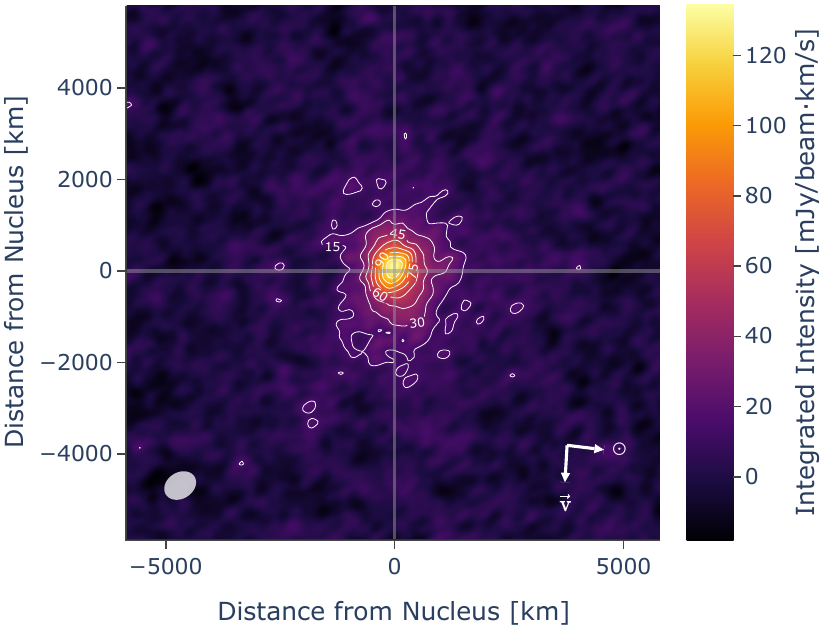}
    \caption{Spectrally integrated intensity map of the \ce{CH3OH} emission lines from a combined ALMA image of the seven March observations. The overlaid contours show the integrated intensity starting at approximately three times the RMS noise, 15 mJy, and incrementing by the same value per level. Arrows in the bottom-right corner indicate the direction of the Sun ($\odot$) and comet velocity ($\vec{\mathrm{v}}$). The ALMA synthesized beam is shown in the bottom-left corner.}
    \label{fig_mom0}
\end{figure}

A moment 0 map of the \ce{CH3OH} emission lines from the ALMA observations is shown in Figure \ref{fig_mom0}. A moment 0 map is a spectrally integrated flux density map, meaning that each pixel shows the total integrated area of the spectrum for all of the \ce{CH3OH} emission lines. The map was generated from the combination of all seven of the March observations, omitting the observation made in April as E3 was too faint to be detected. The morphology of the moment 0 map is centrally peaked and isotropically decreases radially outward. The parameters of the detected \ce{CH3OH} transitions are summarized in Table \ref{table_lines}. LTE modeling of the ALMA Band 6 observations was carried out with a least-squares fit, and the best-fit model is plotted with the spectrum extracted from the \ce{CH3OH} flux peak (Figure \ref{fig_Aspec}). This model fits for the rotational temperature (T\textsubscript{rot}) and column density of a molecule. For the combined image, the modeled T\textsubscript{rot} is 56.1 $\pm$ 0.8 K and the column density is modeled to be (9.70 $\pm$ 0.04)$\times$10\textsuperscript{17} m\textsuperscript{-2}.

\begin{figure*}
    \centering
    \includegraphics[width=\linewidth]{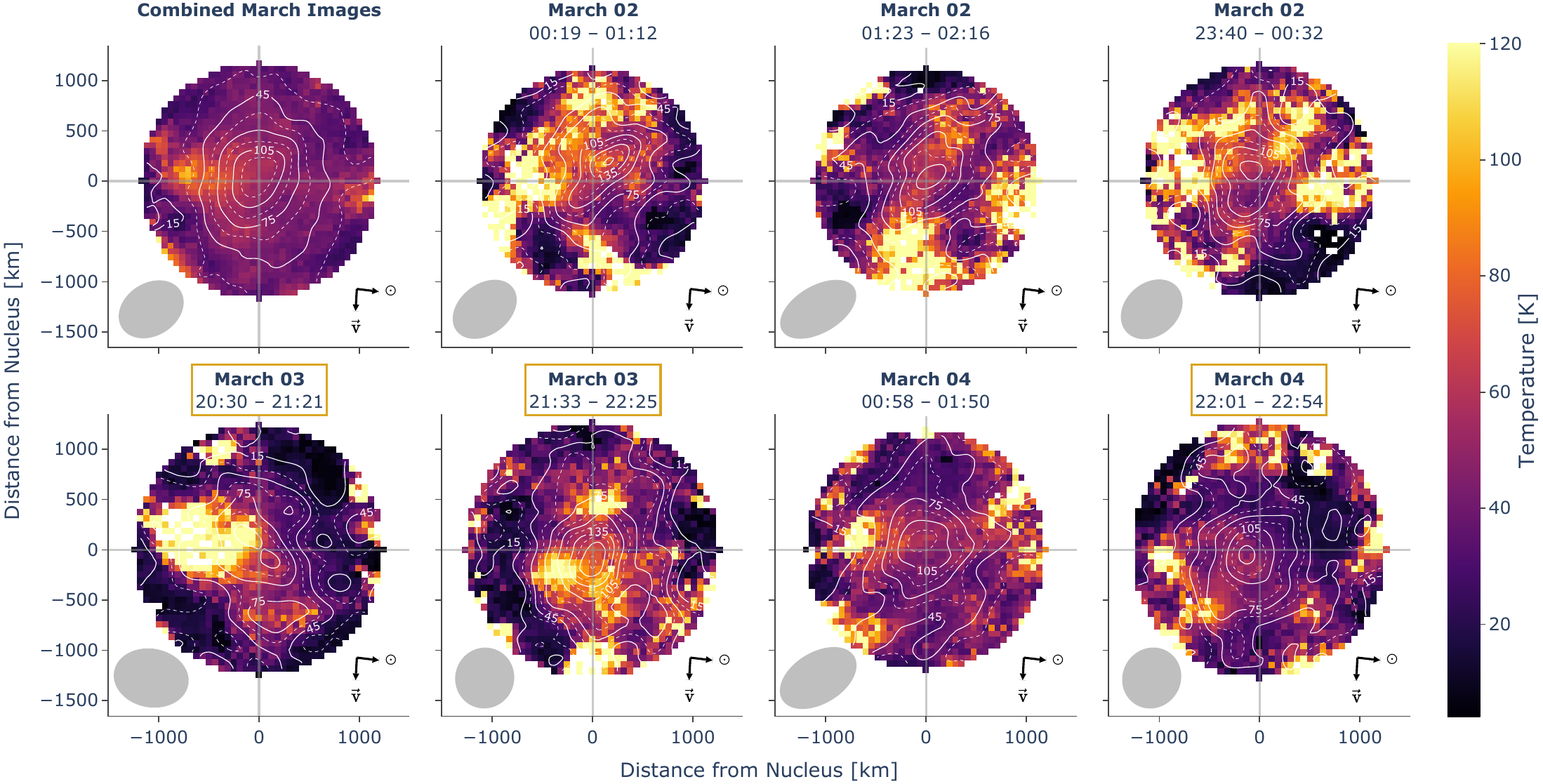}
    \caption{\ce{CH3OH} rotational temperature maps modeled from the seven ALMA observations and the combined image. Integrated intensity maps of the \ce{CH3OH} emission lines for each individual image are overlaid as contours with half-steps (dashed lines) of 15 mJy/beam $\cdot$ km/s. Observations that were measured during the day are indicated by yellow label borders. Arrows in the bottom-right corner indicate the direction of the Sun ($\odot$) and comet velocity ($\vec{\mathrm{v}}$). The ALMA synthesized  beam for each image is shown in the bottom-left corner.}
    \label{fig_ATmaps}
\end{figure*}

\begin{figure*}
    \centering
    \includegraphics[width=\linewidth]{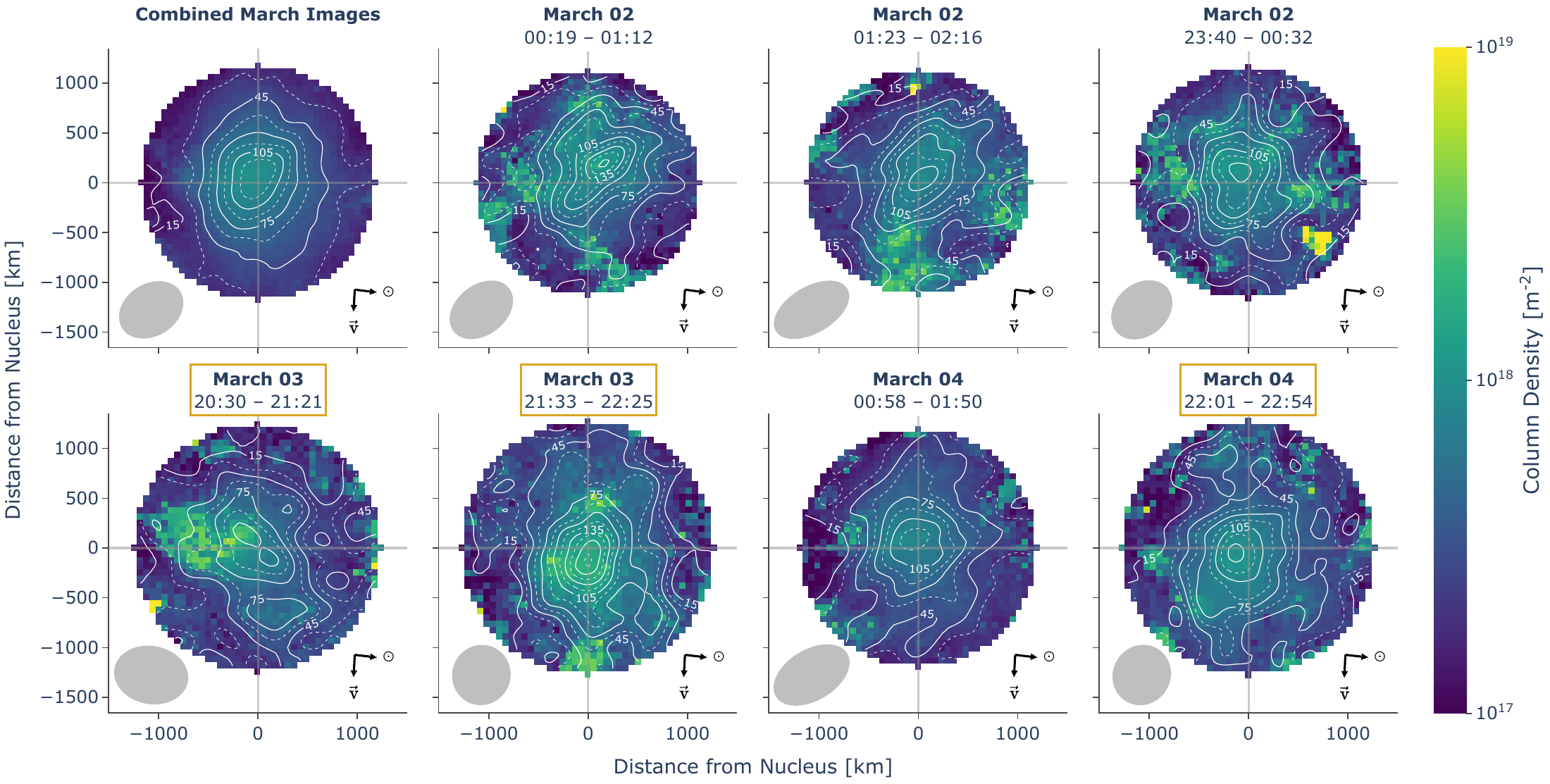}
    \caption{\ce{CH3OH} column density maps modeled from the seven ALMA observations and the combined image. Integrated intensity maps of the \ce{CH3OH} emission lines for each individual image are overlaid as contours with half-steps (dashed lines) of 15 mJy/beam $\cdot$ km/s. Observations that were measured during the day are indicated by yellow label borders. Arrows in the bottom-right corner indicate the direction of the Sun ($\odot$) and comet velocity ($\vec{\mathrm{v}}$). The ALMA synthesized  beam for each image is shown in the bottom-left corner.}
    \label{fig_ANmaps}
\end{figure*}

In \cite{Biver2024a}, the production rates of \ce{H2O} and \ce{CH3OH} exhibit sinusoidal variations with periods (8.4 hrs and 9.1 hrs respectively) similar to the observed rotational period of E3 (8.7 hrs \citep{Knight2023} and 8.5 hrs \citep{Manzini2023}). To investigate this, each of the March observations were individually modeled alongside the combined image. The modeled rotational temperatures are shown in Figure \ref{fig_ATmaps}, with the respective individual \ce{CH3OH} moment 0 maps overlaid as contours. Unlike the JWST modeled column density maps, these maps generally do not show a centrally-peaked distribution. Rather, the peak temperatures tend to be in the anti-sunward direction from the nucleus. Figure \ref{fig_ANmaps} shows the column density maps resulting for each modeled image, with the respective individual \ce{CH3OH} moment 0 maps overlaid as contours. For the most part, the peaks of the modeled column density and integrated flux maps are co-located. Maps that are not centrally peaked show an increase in modeled column density towards the edges of the map, where the confidence in those values decreases (Figure \ref{fig_ANerrmaps}). The values modeled from the combined ALMA image have significantly smaller errors than the average of the individual ALMA values as a result of the lower spectral signal to noise of the individual observations, but the average individual and combined image modeled values are consistent with each other. While modeled column density for the ALMA Band 6 observations tend to be larger than the modeled column density value from the JWST NIRSpec observations, this is possibly a result of how the modeled spectra were extracted. The ALMA spectra were extracted from a single pixel at the photocenter and are thus beam-averaged, which gives more weight to the center of the beam than the edges, resulting in a spectrum that is more sensitive to the areas with more \ce{CH3OH} flux. Conversely, the JWST spectra were extracted in a circular aperture with a radius of 0.47$^{\prime\prime}$ and summed, giving equal weight to each pixel, resulting in uniform sensitivity throughout the aperture. Despite this, the production rates of \ce{CH3OH} are consistent between both observations.

\begin{figure}
    \centering
    \includegraphics[width=\columnwidth]{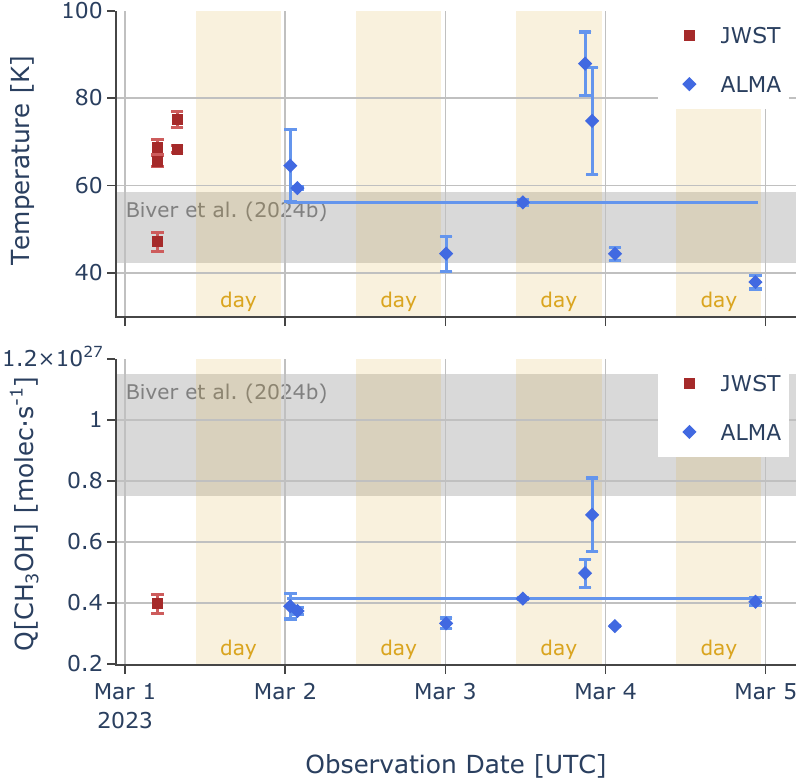}
    \caption{Excitation temperatures (top) and \ce{CH3OH} production rates (bottom) modeled from the JWST (red squares) and ALMA (blue diamonds) observations plotted against time. {The y-error bars are the least-squared errors from the model fits.} The values modeled from the combined ALMA image are represented by a point with a representative time range that covers the time between the first and last March ALMA observations. The shaded yellow regions indicate times between sunrise and sunset and the shaded gray regions indicate the ranges of reported values in \cite{Biver2024a}.}
    \label{fig_TQvt}
\end{figure}

The physical parameters that were modeled in both JWST and ALMA observations are shown with 1$\sigma$ errors in Table \ref{table_physical}. The top section contains the values modeled from 0.47$^{\prime\prime}$-radii circular apertures from the JWST observations, while the middle section has beam-averaged values modeled from the ALMA observations. The bottom section consists of the average of the ALMA observation section, as well as the overall average for values modeled from \ce{H2O} and \ce{CH3OH}. As the ALMA observations were modeled both as a combined image and individually, both the modeled rotational temperature and Q[\ce{CH3OH}] can be plotted against time (Figure \ref{fig_TQvt}) to investigate variations on hour timescales. 

\section{DISCUSSION} \label{discussion}

C/2022 E3 (ZTF) is the first comet to have contemporaneous observations made using both JWST and ALMA, and modeled results show general agreement between the two telescopes. 

Combining the JWST 0.47$^{\prime\prime}$-radius (445 km) circular aperture values and the ALMA resolving beam-averaged values ($\sim$500{\textendash}700 km), the overall average modeled \ce{CH3OH} rotational temperature for E3 during both sets of observations is 57.5 $\pm$ 17.2 K. This is in general agreement with the average excitation temperature modeled from the JWST \ce{H2O} spectral regions at 69.4 $\pm$ 2.9 K. However, the large error on the \ce{CH3OH} average temperature is a result of the two observations taken in succession on March 3, 2023 during the day. The modeled temperatures for these observations have not only large errors, but temperatures higher than the mean at 87.9 $\pm$ 7.3 K and 74.8 $\pm$ 12.2 K, respectively. The \ce{CH3OH} rotational temperature modeled from the ALMA image of all of the March observations has less uncertainty at 56.1 $\pm$ 0.8 K, which is in agreement with the JWST excitation temperature modeled from the \ce{CH3OH} in spectral region [d] at 47.1 $\pm$ 2.2 K. Both of these modeled \ce{CH3OH} excitation temperatures are lower than any individual or average \ce{H2O} excitation temperature. As these are excitation temperatures and not measures of the true kinetic temperature, this disagreement may be a result of non-thermal excitation, or a result of the size of the aperture diluting the \ce{CH3OH} flux density.
Additionally, the overall average \ce{CH3OH} rotational temperature is consistent with previous modeled temperatures from \cite{Biver2024a}, derived from their observations with IRAM on February 4{\textendash}5 2023, when E3 was at a heliocentric distance of 1.17 au. \cite{Biver2024a} reports rotational temperatures derived from \ce{CH3OH} between 42.2 and 58.4 K, from which they derive kinetic temperatures ranging from 54 to 63 K, encompassing nearly all of the temperatures modeled in this work.

\begin{figure*}
    \centering
    \includegraphics[width=\linewidth]{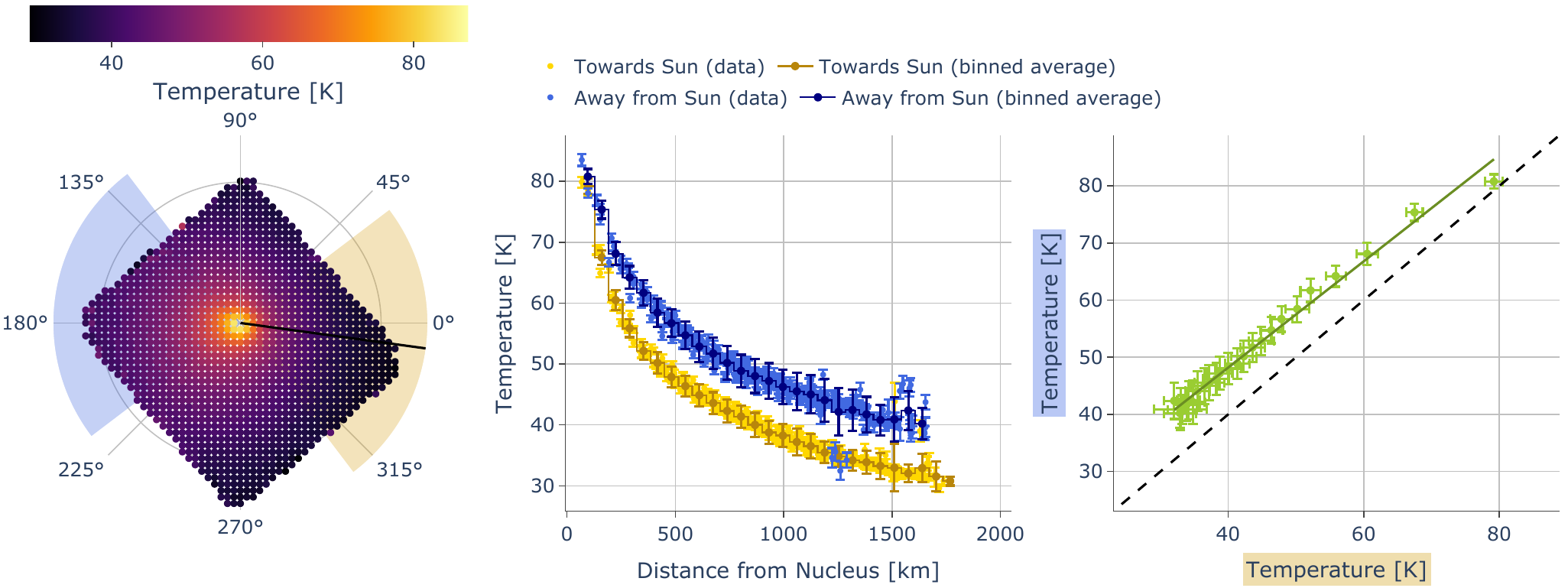}
    \caption{(left) Modeled temperature map for JWST region [a] in Figure \ref{fig_Jspec} with regions within ± 45º of both the Sunward (yellow) and anti-Sunward (blue) vectors. (center) Distributions of the modeled temperatures for the Sun- and anti-Sun regions versus radial distance with bin-averaged distributions overlaid. (right) Bin-averaged modeled temperatures for the away from Sun region (blue) against the towards Sun region (yellow). The linear fit has an equation of $y = (0.930 \pm 0.0216) \cdot x + (1{1.0} \pm 0.9{77})$ and an $\mathrm{R}^2 = 0.98{9}$.}
    \label{fig_antisun-sun}
\end{figure*}

The temperature enhancement in the anti-sunward direction was a surprising result found in the modeled JWST and ALMA temperature maps (Figure \ref{fig_ATmaps}). In order to determine the statistical significance of this enhancement, the temperatures with respect to distance from the nucleus were compared for the sunward and anti-sunward directions. The left subplot in Figure \ref{fig_antisun-sun} shows the temperature map for JWST spectral region [a] with shaded regions indicating if points are within $\pm$ 45$^\circ$ of the plane-of-sky Sunward or anti-Sunward vectors. The Solar phase angle (Sun-Target-Observer) for this observation is 47.8$^\circ$, meaning that roughly the right-hand three-quarters of the visible hemisphere is illuminated by the Sun. In the central subplot of Figure \ref{fig_antisun-sun}, modeled temperatures are plotted as a function of radial distance from the nucleus for the points that fell within $\pm$ 45$^\circ$ of either vector. These points were then bin-averaged into 30 evenly-spaced bins with widths of 64.2 km, similar to the size of a pixel. The temperatures for each averaged bin were plotted against each other in the right subplot in Figure \ref{fig_antisun-sun}. The function y = x has been plotted to show where the points would fall if the modeled temperatures for the Sunward and anti-Sunward regions were equivalent. Since the majority of the points all fall above the y = x line, this indicates that there is an enhancement of temperature in the anti-Sunward direction. As the slopes of most of these graphs are statistically indistinguishable from unity, the statistical significance of this enhancement is determined by the statistical significance of the y-intercept, which is very significant, with a p-value on the order of 10\textsuperscript{-10}. This is the probability that the y-intercept is 0 and means that the enhancement on the anti-sunward side is significant. This enhancement only appears for temperature and is not seen for parameters such as the \ce{CH3OH} column density (see Figure \ref{fig_antisun-sun-Nch3oh}). A similar temperature enhancement was seen in comet 46P and was explained as a result of diminished adiabatic cooling in the anti-Sunward hemisphere, caused by the lower expansion velocity \citep{Cordiner2023}.

\begin{figure}
    \centering
    \includegraphics[width=\columnwidth]{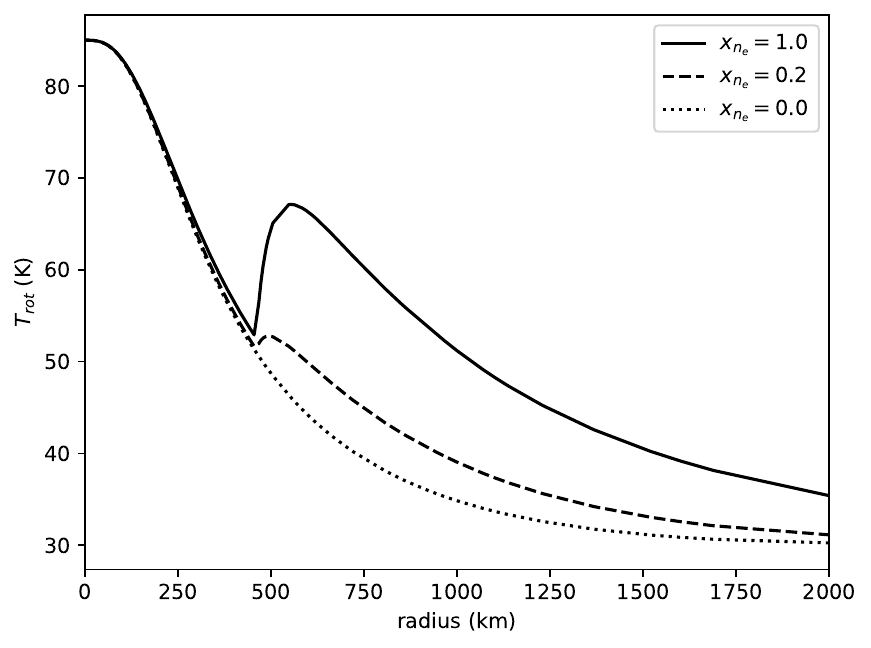}
    \caption{H$_2$O rotational temperature as a function of nucleocentric distance, for a constant coma kinetic temperature of 85 K with varying assumptions for the electron density scaling factor, $x_{ne}$. Based on the SUBLIME radiative transfer model \citep{Cordiner2022,Cordiner2025}.  \label{fig:sublime}}
\end{figure}

The interpretation of rotational temperatures ($T_{rot}$) in cometary comae (particularly at the relatively low densities found at distances $\gtrsim1000$~km from the nucleus), is nontrivial as a result of the breakdown of local thermodynamic equilibrium (LTE) conditions, which means the measured $T_{rot}$ values no longer provide a good measure of the coma kinetic temperature ($T_{kin}$) \citep{bodewits24}. The discrepancy between $T_{rot}$ and $T_{kin}$ can become larger with distance from the nucleus as the gases exit from the collisionally-dominated inner coma, and enter into a radiatively-dominated excitation regime, where collisions with hot electrons produced from H$_2$O photoionization can also play a role.  To shed light on the behavior of the H$_2$O rotational temperatures measured using our JWST data, molecular excitation modeling was performed using the SUBLIME radiative transfer code \citep{Cordiner2022}, which was recently applied to the modeling of cometary H$_2$O by \citet{Cordiner2025}. By solving the balance of collisional and radiative processes as a function of time, in a spherically expanding coma with {$Q({\rm H_2O})=300\times10^{26}$~s$^{-1}$}, and an outflow velocity of 0.9~km\,s$^{-1}$, the H$_2$O energy level populations were calculated as a function of nucleocentric distance ($r$). The model $T_{rot}(r)$ curves (Figure \ref{fig:sublime}) were then derived by fitting Boltzmann distributions to the populations of the first nine energy levels of ortho-H$_2$O, as a function of radius. A constant kinetic temperature of 85~K was used, and the electron density scaling factor ($x_{ne}$) was varied between 0.0 to 1.0 to investigate the importance of electron collisions in comet E3. 

As shown by comparison of the model $T_{rot}(r)$ results in Figure \ref{fig:sublime} with the observed $T_{rot}(r)$ curve in Figure \ref{fig_antisun-sun}, the decreasing H$_2$O rotational temperatures with distance in comet E3's coma can be explained largely as a result of non-LTE molecular excitation, in the presence of a constant kinetic temperature of $\sim85$~K. The steady decrease of the model $T_{rot}(r)$ curve in the case of $x_{ne}=0$ is primarily a result of rotational line cooling of H$_2$O, which leads to a downward cascade of the rotational level populations towards lower energy, as the gas flows outward through the coma. There are insufficient collisions within the neutral gas to maintain LTE in this region. However, in the presence of hot electrons ($x_{ne}=1.0$) there is a significant increase in the modeled H$_2$O rotational temperatures around the radius of the contact surface  \citep[see \emph{e.g.}][]{biver99}, due to a rapid increase in the electron densities and temperatures in this region. Typically, an electron density scaling factor $x_{ne}=0.2$ is employed for modeling H$_2$O in cometary comae (see \citealt{Cordiner2025} and references therein), and in that case, the impact of electron collisions on $T_{rot}$ is more moderate (Figure \ref{fig:sublime}, solid trace). Ultimately, the parameterization of the electron densities and temperatures adopted by \citet{biver99} and \citet{Cordiner2025} likely represents a relatively crude, analytic approximation to the true coma conditions, which are expected to vary from comet to comet depending on the specific coma ionization state and any solar wind interactions that may accelerate or disperse the electrons.  Therefore, the lack of a significant bump in the observed $T_{rot}$ values around the contact surface ($r\sim750$~km) should not be taken as strong evidence for a lack of hot electrons, but could imply that the onset of electron collisional heating is smoother than implied by the \citet{biver99} parameterization. Additional, detailed magnetohydrodynamic and radiative transfer modeling of the electron distribution in comet E3 will be required to shed more light on this issue, to help further investigate whether the H$_2$O rotational temperatures as a function of radius may provide a useful probe of the coma ionization state and electron temperatures, in addition to the kinetic temperature of the neutral coma.

At the center of the coma, the average modeled value for the \ce{H2O} production rate is {(309 $\pm$ 11)$\times$10\textsuperscript{26}} molec.~s\textsuperscript{-1}. This is lower than the Q[\ce{H2O}] measured for E3 with Odin in UT January 2023, when E3 had an r\textsubscript{h} of 1.12 au, in \cite{Biver2024a}, which ranges from {$\sim$500$\times$10\textsuperscript{26}} to {$\sim$800$\times$10\textsuperscript{26}} molec.~s\textsuperscript{-1}. {While there is a general agreement in the \ce{H2O} production rate for the two observations, the value decreases by approximately 30\% from the first to the second. This appears to be consistent with the temporal variations observed in both \ce{H2O} and \ce{CH3OH} by \cite{Biver2024a}, where the production rates of \ce{H2O} varied by $\pm$5\%. This variability appears sinusoidal, with a period of 8.4 hours, which is approximately the determined rotation period of the comet \citep{Knight2023, Manzini2023}. Given that the two JWST observations were about 3 hours apart, the decrease in the \ce{H2O} production rate may be a continuation of this sinusoidal trend and the higher degree of variability may be due to the higher spatial resolution of JWST NIRSpec as compared to the Odin observations in \cite{Biver2024a}.}

The average Q[\ce{CH3OH}] from both JWST and ALMA is (4.26 $\pm$ 1.41)$\times$10\textsuperscript{26} molec.~s\textsuperscript{-1}, which results in a relative abundance of \ce{CH3OH} of 1.38 $\pm$ 0.46\%. While this abundance is within the range of previous \ce{CH3OH} detections in other comets, it is indicative of a methanol-poor comet \citep{Mumma2011, Biver2024b}. The range of \ce{CH3OH} production rates calculated in this work are all lower than the Q[\ce{CH3OH}] reported in \cite{Biver2024a} of (8.78 $\pm$ 0.08)$\times$10\textsuperscript{26} molec.~s\textsuperscript{-1}{, as is shown in Figure \ref{fig_TQvt}}. As the heliocentric distance of E3 increased by $\sim$0.18 au between the \cite{Biver2024a} observations and those in this work, the production rates would be expected to either remain constant or decrease slightly, so the values presented in this work appear to be consistent. {In Figure \ref{fig_TQvt}, there are two observations that not only have higher modeled temperatures and \ce{CH3OH} production rates, but also larger errors. While the variability in these observations could be real, perhaps due to a short-lived outburst of \ce{CH3OH}-rich material, it is hard to rule out calibration artefacts as the source of the variability. As such, these observations have been included for completeness.} Finally, the \ce{CH3OH} abundance reported in \cite{Biver2024a} of 1.76 $\pm$ 0.01\% wrt \ce{H2O} is within the error of the value found in this work. For comet observations with heliocentric distances less than 2 au, the relative abundances of molecules tends to stay constant \citep{Lippi2021}, indicating that the rates of decline for both Q[\ce{H2O}] and Q[\ce{CH3OH}] should be similar. Although the modeled values from this work are consistent with the derived values in \cite{Biver2024a}, the IRAM observations, as single-dish observations, do not have the mapping capabilities to show where the bulk of the emission is located, as well as the asymmetric temperature distribution of the coma. {The asymmetry of the temperature distribution may help explain why the modeled temperatures in Figure \ref{fig_TQvt} do not show a decrease between the values reported in \cite{Biver2024a} and the values in this paper, while the \ce{CH3OH} production rates do.}

Figure \ref{fig_Jact} shows the modeled local gas production rate maps for the four \ce{H2O} spectral regions. The top two are from the first NIRSpec observation, and the bottom two are from the second, observed around 2 hours later. All four maps show a decrease in the center from opacity and PSF effects, which extends away from the center in the southeast direction ($\sim$135$^\circ$ clockwise from North) for the maps in the top row, and in the northeast direction ($\sim$45$^\circ$ clockwise from North) for the bottom row. These show a rotation of the comet in the plane of the observations of $\sim$90$^{\circ}$ in 2 hours, indicating a full rotation time of $\sim$8 hours, which is comparable to the rotational period derived from earlier observations \citep{Knight2023, Manzini2023}.

In Figure \ref{fig_Aspec}, the nearly all of the emission is coming from \ce{CH3OH}. However, at 241.7743 GHz, there is an emission spike with a detection level of 2.5$\sigma${, which is highlighted in green in Figure \ref{fig_Aspec}}. There is one molecule that has been previously been detected in comets, HNCO, which has its $J=11_{0,11}-10_{0,10}$ line at 241.774032 GHz. As this is the only line of HNCO that is detected and the signal-to-noise ratio is less than 3, the unambiguous detection of this line is questionable, though it's likelihood is aided by the detection of this same line in the \cite{Biver2024a} IRAM observations. The peak of this emission is not the same as that of \ce{CH3OH}, rather it is centered on a pixel directly southeast of the \ce{CH3OH} peak (approximately 83 km from the nucleus). Assuming that HNCO has the same rotational temperature as \ce{CH3OH}, the column density calculated at this pixel is (2.25 $\pm$ 0.70)$\times$10\textsuperscript{16} m\textsuperscript{-2}. The production rate of HNCO is calculated to be (9.63 $\pm$ 3.00)$\times$10\textsuperscript{24} molec.~s\textsuperscript{-1}, which corresponds to a molecular abundance of 0.031 $\pm$ 0.010\%, both of which are around half of the values reported in \cite{Biver2024a}. HNCO also absorbs in the infrared \citep{Sharpe2004, Kochanov2019, Gordon2022}, though at the abundance reported in \cite{Biver2024a}, the lines are not only too faint to detect in the JWST NIRSpec observations, but are also at wavelengths in regions dominated by \ce{H2O} emission \citep{Villanueva2012a}.

JWST spectral region [e] (4.50{\textendash}5.23 µm), while dominated by \ce{H2O} emission, also contains emission from \ce{CO}, \ce{CN}, and \ce{OCS}. \ce{OCS} is one of the few S-bearing species that is detectable in the infrared, as the \ce{C-O} stretch ($\nu_3$-band) emits strongly at 4.85 µm \citep{DelloRusso1998, Saki2020}. While sulfur is one of the most common elements in the interstellar medium, observations of star- and planetary system-forming regions find abundances of sulfur-bearing molecules to be orders of magnitude lower than expected \citep{Calmonte2016}. The same holds true for comets{\textemdash}as of 2022, despite the strength of the vibrational transition, \ce{OCS} has only been detected in 10 comets \citep{Saki2020, Biver2024b}{, including 67P/Churyumov-Gerasimenko \citep{Lauter2020}}. There are several hypotheses for the reservoir of missing sulfur, from S-bearing molecules being frozen onto icy grains \citep{Saki2020} to a majority of the sulfur residing in sulfur allotropes, which are generally stable rings like \ce{S4} or \ce{S8} \citep{Calmonte2016}. As for the few detections of \ce{OCS}, \cite{Saki2020} suggests that it is under-represented in cometary studies and thus represents a bias in observations rather than an actual phenomena.

\begin{figure*}
    \centering
    \includegraphics[width=0.8\linewidth]{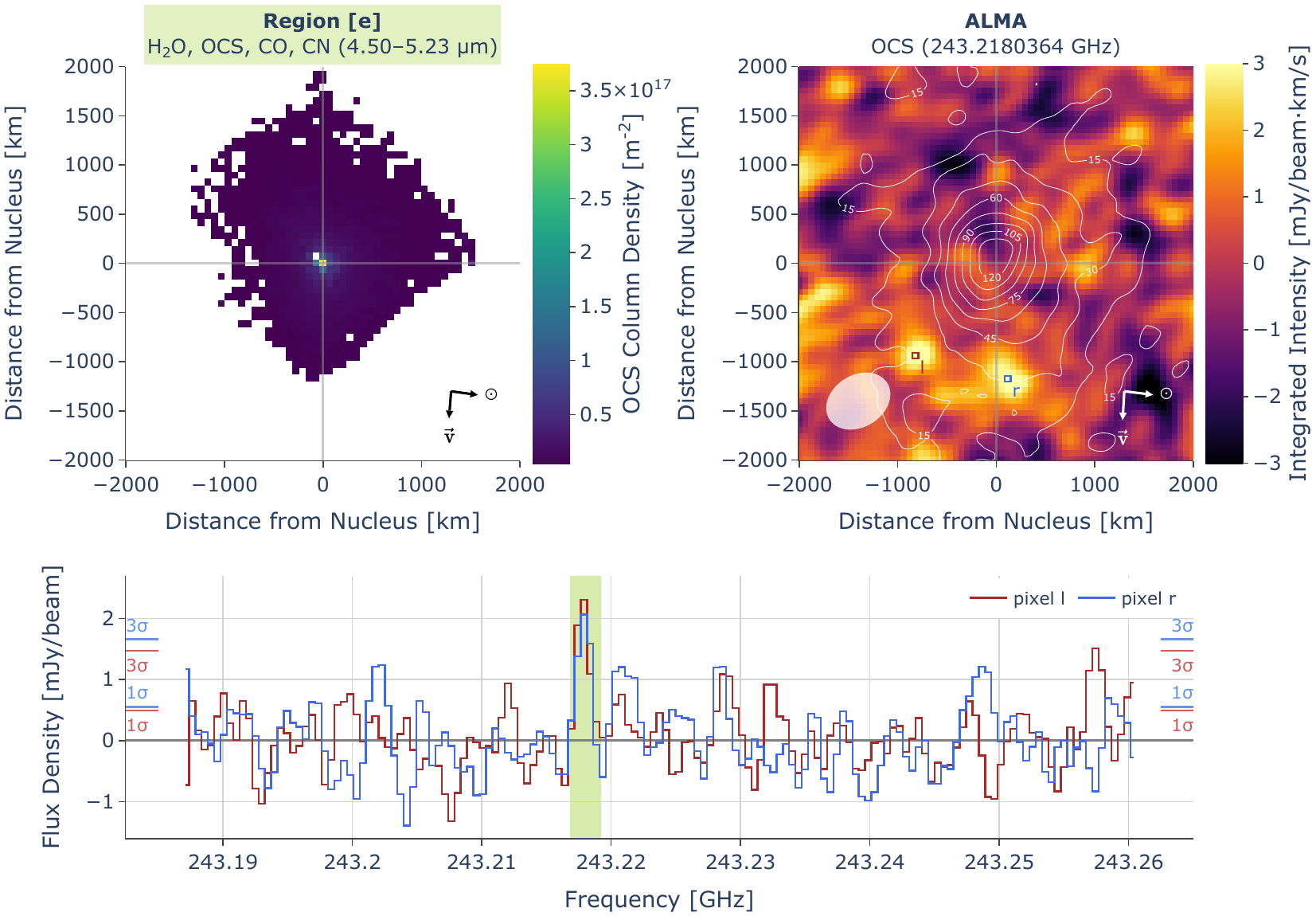}
    \caption{(left) Modeled column density of OCS from JWST spectral region [e]. (right) Moment 0 map of the \ce{OCS} $J=20-19$ emission line from the combined image of all of the ALMA March observations. Contours show the \ce{CH3OH} moment 0 map from the same image cube (Figure \ref{fig_mom0}). The OCS peaks are indicated by two squares (l, red; r, blue).  (bottom) Spectra extracted from the cube for the two pixel in the moment map. The respective 1 and 3 RMS noise levels are indicated in the margins and the highlighted green region shows the frequency of the \ce{OCS} line.}
    \label{fig_alma-ocs}
\end{figure*}

Similarly to \ce{CH3OH}, \ce{OCS} presents an opportunity for a multi-wavelength investigation with JWST and ALMA as, in addition to the 4.85 µm emission, it has several rotational lines that may be detectable by ALMA \citep{Endres2016}. For the 0.47$^{\prime\prime}$ circular aperture and using the \ce{H2O} production rate from spectral region [e], the modeled molecular abundance, production rate, and column density of OCS are 0.044 $\pm$ 0.003\% wrt \ce{H2O}, (1.58 $\pm$ 0.10)$\times$10\textsuperscript{25} molec.~s\textsuperscript{-1}, and (1.85 $\pm$ 0.12)$\times$10\textsuperscript{16} m\textsuperscript{-2} respectively. Shown in the top-left subplot of Figure \ref{fig_alma-ocs} is a map of the modeled column density from the JWST NIRSpec observation of \ce{OCS} (region [e]). However, as the \ce{OCS} lines that are detectable by ALMA are both fewer and sparser than in the infrared, in the observational set-up for these observations of E3, only the \ce{OCS} $J=20-19$ line is covered by any of the ALMA spectral windows. The top-right subplot of Figure \ref{fig_alma-ocs} shows the moment 0 map of the $J=20-19$ emission line with the \ce{CH3OH} moment 0 map overlaid as contours. The \ce{OCS} moment map was convolved with a 1.5$^{\prime\prime}$ circular beam to improve the signal-to-noise ratio. The OCS emission has two peaks, which are indicated on the moment 0 map as color-coded squares. The spectra extracted from the two \ce{OCS} emission peaks are shown in the bottom panel of Figure \ref{fig_alma-ocs}, where the \ce{OCS} line is detected at 4.7$\sigma$ for left (red) peak and 3.7$\sigma$ for the right (blue) peak. Given that only one line was observed at low significance and, in the ALMA observations, not observed co-spatially with \ce{CH3OH} or even the peak of the continuum phase-center, the \ce{OCS} detection in the ALMA observations is tentative. If the rotational temperature of OCS is assumed to be the same as that of \ce{CH3OH}, the column density for the two pixels are determined to be very similar, with (1.29 $\pm$ 0.28)$\times$10\textsuperscript{17} m\textsuperscript{-2} for the left pixel and (1.20 $\pm$ 0.27)$\times$10\textsuperscript{17} m\textsuperscript{-2} for the right pixel. The production rates are determined to be (5.55 $\pm$ 1.19)$\times$10\textsuperscript{25} molec.~s\textsuperscript{-1} for the left pixel and (5.18 $\pm$ 1.15)$\times$10\textsuperscript{25} molec.~s\textsuperscript{-1} for the right pixel. Using the average \ce{H2O} production rate, these production rates correspond to molecular abundances of 0.180 $\pm$ 0.039\% and 0.168 $\pm$ 0.038\% for the left and right pixels respectively. While the modeled JWST production rate and molecular abundance are lower than those reported in \cite{Biver2024a}, the values from both of the ALMA modeled pixels are higher, likely as a result of the assumed temperature and lack of other lines. Further discussion of the JWST observations of \ce{OCS} will be in S. Milam et al. (in prep.).

More robust ALMA observations of cometary \ce{OCS} would have the benefit of measuring a less contaminated spectrum than JWST, as \ce{OCS} emits in the infrared at similar wavelengths to several other strong emitters like \ce{H2O}, \ce{CO}, and \ce{CN}. As a result, whereas models of JWST observations determine a combined excitation temperature for all of the molecules in the wavelength region, models of ALMA observations would determine a rotational temperature specific to \ce{OCS}. Future spectral studies of comets that utilize contemporaneous observations with both JWST and ALMA should aim to target several molecules that emit in both wavelength regimes, especially molecules that are historically under-observed, like \ce{OCS}.

\section{CONCLUSIONS} \label{conclusions}

For the first time, contemporaneous  spectroscopic observations with both JWST and ALMA were made of a comet, C/2022 E3 (ZTF) in March of 2023. To investigate the similarities between derived values, the spectra from both telescopes were modeled to derive the same parameters: rotational temperatures and molecular column densities. This study focused on \ce{CH3OH} as it was detected with both JWST and ALMA.

For the modeled \ce{CH3OH} T\textsubscript{rot} and N[\ce{CH3OH}], as well as the calculated Q[\ce{CH3OH}], the JWST values were within the error bars of the ALMA average values, indicating general agreement between both telescopes and modeling methods. Additionally, E3 displays a statistically significant enhancement in modeled temperature on the anti-sunward side of the coma that is not evident in \ce{CH3OH} column density.

According to our radiative transfer modeling using the SUBLIME code, the decreasing \ce{H2O} rotational temperatures with distance in comet E3's coma can be explained largely as a result of non-LTE molecular excitation, in the presence of a constant kinetic temperature of $\sim85$~K.  The presence of a strong increase in rotational temperatures around the radius of the contact surface ($r\sim750$~km) was not observed, which may imply that electron collisional heating was less important in comet E3, or that the onset of electron collisional heating occurs more smoothly in E3 than typically assumed.

The modeled values for both JWST and ALMA are also consistent with results from earlier single-dish observations of E3 presented in \cite{Biver2024a}. The production rates of both \ce{H2O} and \ce{CH3OH} appear to have decreased slightly, but this was expected due to the increase in heliocentric distance between the observations.

These results emphasize the advantages of performing contemporaneous spectroscopic observations, particularly those that sample different parts of the electromagnetic spectrum. With observatories like ALMA accepting joint proposals to coordinate observing on multiple telescopes, more comets can be studied with contemporaneous observations. These observations will increase the population of comets with well-studied chemical compositions.

\begin{acknowledgments}
The National Radio Astronomy Observatory is a facility of the National Science Foundation operated under cooperative agreement by Associated Universities, Inc.  This paper makes use of the following ALMA data: ADS/JAO.ALMA\#2022.1.00997.T. ALMA is a partnership of ESO (representing its member states), NSF (USA) and NINS (Japan), together with NRC (Canada), MOST and ASIAA (Taiwan), and KASI (Republic of Korea), in cooperation with the Republic of Chile. The Joint ALMA Observatory is operated by ESO, AUI/NRAO and NAOJ.

This work is based in part on observations made with the NASA/ESA/CSA James Webb Space Telescope. The data were obtained from the Mikulski Archive for Space Telescopes at the Space Telescope Science Institute, which is operated by the Association of Universities for Research in Astronomy, Inc., under NASA contract NAS 5-03127 for JWST. These observations are associated with program \#1253 and can be accessed at \dataset[https://doi.org/10.17909/czbe-fk30]{https://doi.org/10.17909/czbe-fk30}.

SNM, NXR, and MAC were supported by the Planetary Science Division Internal Scientist Funding Program through the Fundamental Laboratory Research (FLaRe) work package. SNM also acknowledges support from NASA JWST Interdisciplinary Scientist grant 21-SMDSS21-0013.

Part of this research was carried out at the Jet Propulsion Laboratory, California Institute of Technology, under a contract with the National Aeronautics and Space Administration (80NM0018D0004). D.C.L. acknowledges financial support from the National Aeronautics and Space Administration (NASA) Astrophysics Data Analysis Program (ADAP).

\end{acknowledgments}

\bibliography{research}{}

@ARTICLE{Lauter2020,
       author = {{L{\"a}uter}, Matthias and {Kramer}, Tobias and {Rubin}, Martin and {Altwegg}, Kathrin},
        title = "{The gas production of 14 species from comet 67P/Churyumov-Gerasimenko based on DFMS/COPS data from 2014 to 2016}",
      journal = {Monthly Notices of the Royal Astronomical Society},
         year = 2020,
        month = nov,
       volume = {498},
       number = {3},
        pages = {3995-4004},
          doi = {10.1093/mnras/staa2643},
archivePrefix = {arXiv},
       eprint = {2006.01750},
 primaryClass = {astro-ph.EP},
       adsurl = {https://ui.adsabs.harvard.edu/abs/2020MNRAS.498.3995L}
}

@ARTICLE{Taylor2017,
       author = {{Taylor}, M.~G.~G.~T. and {Altobelli}, N. and {Buratti}, B.~J. and {Choukroun}, M.},
        title = "{The Rosetta mission orbiter science overview: the comet phase}",
      journal = {Philosophical Transactions of the Royal Society of London Series A},
         year = 2017,
        month = may,
       volume = {375},
       number = {2097},
          eid = {20160262},
        pages = {20160262},
          doi = {10.1098/rsta.2016.0262},
archivePrefix = {arXiv},
       eprint = {1703.10462},
 primaryClass = {astro-ph.EP},
       adsurl = {https://ui.adsabs.harvard.edu/abs/2017RSPTA.37560262T}
}

@ARTICLE{Rubin2019,
       author = {{Rubin}, Martin and {Altwegg}, Kathrin and {Balsiger}, Hans and {Berthelier}, Jean-Jacques and {Combi}, Michael R. and {De Keyser}, Johan and {Drozdovskaya}, Maria and {Fiethe}, Bj{\"o}rn and {Fuselier}, Stephen A. and {Gasc}, S{\'e}bastien and {Gombosi}, Tamas I. and {H{\"a}nni}, Nora and {Hansen}, Kenneth C. and {Mall}, Urs and {R{\`e}me}, Henri and {Schroeder}, Isaac R.~H.~G. and {Schuhmann}, Markus and {S{\'e}mon}, Thierry and {Waite}, Jack H. and {Wampfler}, Susanne F. and {Wurz}, Peter},
        title = "{Elemental and molecular abundances in comet 67P/Churyumov-Gerasimenko}",
      journal = {Monthly Notices of the Royal Astronomical Society},
         year = 2019,
        month = oct,
       volume = {489},
       number = {1},
        pages = {594-607},
          doi = {10.1093/mnras/stz2086},
archivePrefix = {arXiv},
       eprint = {1907.11044},
 primaryClass = {astro-ph.EP},
       adsurl = {https://ui.adsabs.harvard.edu/abs/2019MNRAS.489..594R}
}

@ARTICLE{Filacchione2019,
       author = {{Filacchione}, Gianrico and {Groussin}, Olivier and {Herny}, Cl{\'e}mence and {Kappel}, David and {Mottola}, Stefano and {Oklay}, Nilda and {Pommerol}, Antoine and {Wright}, Ian and {Yoldi}, Zurine and {Ciarniello}, Mauro and {Moroz}, Lyuba and {Raponi}, Andrea},
        title = "{Comet 67P/CG Nucleus Composition and Comparison to Other Comets}",
      journal = {Space Science Reviews},
         year = 2019,
        month = feb,
       volume = {215},
       number = {1},
          eid = {19},
        pages = {19},
          doi = {10.1007/s11214-019-0580-3},
       adsurl = {https://ui.adsabs.harvard.edu/abs/2019SSRv..215...19F}
}

@ARTICLE{Keller2020,
       author = {{Keller}, Horst Uwe and {K{\"u}hrt}, Ekkehard},
        title = "{Cometary Nuclei{\textemdash}From Giotto to Rosetta}",
      journal = {Space Science Reviews},
         year = 2020,
        month = jan,
       volume = {216},
       number = {1},
          eid = {14},
        pages = {14},
          doi = {10.1007/s11214-020-0634-6},
       adsurl = {https://ui.adsabs.harvard.edu/abs/2020SSRv..216...14K}
}

@ARTICLE{Cordiner2025,
       author = {{Cordiner}, M.~A. and {Gibb}, E.~L. and {Kisiel}, Z. and {Roth}, N.~X. and {Biver}, N. and {Bockel{\'e}e-Morvan}, D. and {Boissier}, J. and {Bonev}, B.~P. and {Charnley}, S.~B. and {Coulson}, I.~M. and {Crovisier}, J. and {Drozdovskaya}, M.~N. and {Furuya}, K. and {Jin}, M. and {Kuan}, Y. -J. and {Lippi}, M. and {Lis}, D.~C. and {Milam}, S.~N. and {Opitom}, C. and {Qi}, C. and {Remijan}, A.~J.},
        title = "{A D/H ratio consistent with Earth's water in Halley-type comet 12P from ALMA HDO mapping}",
      journal = {Nature Astronomy},
         year = 2025,
        month = aug,
          doi = {10.1038/s41550-025-02614-7},
       adsurl = {https://ui.adsabs.harvard.edu/abs/2025NatAs.tmp..165C}
}

@misc{mpfit,
    author = {{Koposov}, Sergey and {Rivers}, Mark and {Markwardt}, Craig and {Garbow}, B and {Hillstrom}, K and {More}, J},
  title        = {mpfit},
  year         = 2017,
  publisher    = {Github},
  url          = {https://github.com/segasai/astrolibpy/blob/master/mpfit/mpfit.py}
}

@INCOLLECTION{bodewits24,
       author = {{Bodewits}, D. and {Bonev}, B.~P. and {Cordiner}, M.~A. and {Villanueva}, G.~L.},
        title = "{Radiative Processes as Diagnostics of Cometary Atmospheres}",
        publisher={University of Arizona Press},
    booktitle = {Comets III},
         year = 2024,
       editor = {{Meech}, Karen. J. and {Combi}, Michael. R. and {Bockel{\'e}e-Morvan}, Dominique and {Raymodn}, Sean. N. and {Zolensky}, Michael. E.},
        pages = {407-432},
          doi = {10.2458/azu_uapress_9780816553631-ch013},
       adsurl = {https://ui.adsabs.harvard.edu/abs/2024come.book..407B}
}

@ARTICLE{biver99,
       author = {{Biver}, N. and {Bockel{\'e}e-Morvan}, D. and {Crovisier}, J. and {Davies}, J.~K. and {Matthews}, H.~E. and {Wink}, J.~E. and {Rauer}, H. and {Colom}, P. and {Dent}, W.~R.~F. and {Despois}, D. and {Moreno}, R. and {Paubert}, G. and {Jewitt}, D. and {Senay}, M.},
        title = "{Spectroscopic Monitoring of Comet C/1996 B2 (Hyakutake) with the JCMT and IRAM Radio Telescopes}",
      journal = {\aj},
         year = 1999,
        month = oct,
       volume = {118},
       number = {4},
        pages = {1850-1872},
          doi = {10.1086/301033},
       adsurl = {https://ui.adsabs.harvard.edu/abs/1999AJ....118.1850B}
}

@article{Jakobsen2022,
	author = {{Jakobsen, P.} and {Ferruit, P.} and {Alves de Oliveira, C.} and {Arribas, S.} and {Bagnasco, G.} and {Barho, R.} and {Beck, T. L.} and {Birkmann, S.} and {Böker, T.} and {Bunker, A. J.} and {Charlot, S.} and {de Jong, P.} and {de Marchi, G.} and {Ehrenwinkler, R.} and {Falcolini, M.} and {Fels, R.} and {Franx, M.} and {Franz, D.} and {Funke, M.} and {Giardino, G.} and {Gnata, X.} and {Holota, W.} and {Honnen, K.} and {Jensen, P. L.} and {Jentsch, M.} and {Johnson, T.} and {Jollet, D.} and {Karl, H.} and {Kling, G.} and {Köhler, J.} and {Kolm, M.-G.} and {Kumari, N.} and {Lander, M. E.} and {Lemke, R.} and {López-Caniego, M.} and {Lützgendorf, N.} and {Maiolino, R.} and {Manjavacas, E.} and {Marston, A.} and {Maschmann, M.} and {Maurer, R.} and {Messerschmidt, B.} and {Moseley, S. H.} and {Mosner, P.} and {Mott, D. B.} and {Muzerolle, J.} and {Pirzkal, N.} and {Pittet, J.-F.} and {Plitzke, A.} and {Posselt, W.} and {Rapp, B.} and {Rauscher, B. J.} and {Rawle, T.} and {Rix, H.-W.} and {Rödel, A.} and {Rumler, P.} and {Sabbi, E.} and {Salvignol, J.-C.} and {Schmid, T.} and {Sirianni, M.} and {Smith, C.} and {Strada, P.} and {te Plate, M.} and {Valenti, J.} and {Wettemann, T.} and {Wiehe, T.} and {Wiesmayer, M.} and {Willott, C. J.} and {Wright, R.} and {Zeidler, P.} and {Zincke, C.}},
	title = {The Near-Infrared Spectrograph (NIRSpec) on the James Webb Space Telescope - I. Overview of the instrument and its capabilities},
	DOI= "10.1051/0004-6361/202142663",
	url= "https://doi.org/10.1051/0004-6361/202142663",
	journal = {Astronomy \& Astrophysics},
	year = 2022,
	volume = 661,
	pages = "A80",
}

@ARTICLE{Ehrenfreund2000,
       author = {{Ehrenfreund}, Pascale and {Charnley}, Steven B.},
        title = "{Organic Molecules in the Interstellar Medium, Comets, and Meteorites: A Voyage from Dark Clouds to the Early Earth}",
      journal = {Annual Review of Astronomy and Astrophysics},
         year = 2000,
        month = jan,
       volume = {38},
        pages = {427-483},
          doi = {10.1146/annurev.astro.38.1.427},
       adsurl = {https://ui.adsabs.harvard.edu/abs/2000ARA\&A..38..427E}
}

@ARTICLE{Cordiner2022,
       author = {{Cordiner}, M.~A. and {Coulson}, I.~M. and {Garcia-Berrios}, E. and {Qi}, C. and {Lique}, F. and {Zo{\l}towski}, M. and {de Val-Borro}, M. and {Kuan}, Y. -J. and {Ip}, W. -H. and {Mairs}, S. and {Roth}, N.~X. and {Charnley}, S.~B. and {Milam}, S.~N. and {Tseng}, W. -L. and {Chuang}, Y. -L.},
        title = "{A SUBLIME 3D Model for Cometary Coma Emission: The Hypervolatile-rich Comet C/2016 R2 (PanSTARRS)}",
      journal = {Astrophysical Journal},
         year = 2022,
        month = apr,
       volume = {929},
       number = {1},
          eid = {38},
        pages = {38},
          doi = {10.3847/1538-4357/ac5893},
archivePrefix = {arXiv},
       eprint = {2202.11849},
 primaryClass = {astro-ph.EP},
       adsurl = {https://ui.adsabs.harvard.edu/abs/2022ApJ...929...38C}
}

@ARTICLE{DelloRusso1998,
       author = {{Dello Russo}, Neil and {DiSanti}, Michael A. and {Mumma}, Michael J. and {Magee-Sauer}, Karen and {Rettig}, Terrence W.},
        title = "{Carbonyl Sulfide in Comets C/1996 B2 (Hyakutake) and C/1995 O1 (Hale-Bopp): Evidence for an Extended Source in Hale-Bopp}",
      journal = {Icarus},
         year = 1998,
        month = oct,
       volume = {135},
       number = {2},
        pages = {377-388},
          doi = {10.1006/icar.1998.5990},
       adsurl = {https://ui.adsabs.harvard.edu/abs/1998Icar..135..377D}
}

@ARTICLE{Saki2020,
       author = {{Saki}, Mohammad and {Gibb}, Erika L. and {Bonev}, Boncho P. and {Roth}, Nathan X. and {DiSanti}, Michael A. and {Dello Russo}, Neil and {Vervack}, Jr., Ronald J. and {McKay}, Adam J. and {Kawakita}, Hideyo},
        title = "{Carbonyl Sulfide (OCS): Detections in Comets C/2002 T7 (LINEAR), C/2015 ER61 (PanSTARRS), and 21P/Giacobini-Zinner and Stringent Upper Limits in 46P/Wirtanen}",
      journal = {Astronomical Journal},
         year = 2020,
        month = oct,
       volume = {160},
       number = {4},
          eid = {184},
        pages = {184},
          doi = {10.3847/1538-3881/aba522},
       adsurl = {https://ui.adsabs.harvard.edu/abs/2020AJ....160..184S}
}

@ARTICLE{Calmonte2016,
       author = {{Calmonte}, U. and {Altwegg}, K. and {Balsiger}, H. and {Berthelier}, J.~J. and {Bieler}, A. and {Cessateur}, G. and {Dhooghe}, F. and {van Dishoeck}, E.~F. and {Fiethe}, B. and {Fuselier}, S.~A. and {Gasc}, S. and {Gombosi}, T.~I. and {H{\"a}ssig}, M. and {Le Roy}, L. and {Rubin}, M. and {S{\'e}mon}, T. and {Tzou}, C. -Y. and {Wampfler}, S.~F.},
        title = "{Sulphur-bearing species in the coma of comet 67P/Churyumov-Gerasimenko}",
      journal = {Monthly Notices of the Royal Astronomical Society},
         year = 2016,
        month = nov,
       volume = {462},
        pages = {S253-S273},
          doi = {10.1093/mnras/stw2601},
       adsurl = {https://ui.adsabs.harvard.edu/abs/2016MNRAS.462S.253C}
}

@article{Gordon2022,
author = {I.~E. {Gordon} AND L.~S. {Rothman} AND R.~J. {Hargreaves} AND R. {Hashemi} AND E.~V. {Karlovets} AND F.~M. {Skinner} AND E.~K. {Conway} AND C. {Hill} AND R.~V. {Kochanov} AND Y. {Tan} AND P. {Wcis{\l}o} AND A.~A. {Finenko} AND K. {Nelson} AND P.~F. {Bernath} AND M. {Birk} AND V. {Boudon} AND A. {Campargue} AND K.~V. {Chance} AND A. {Coustenis} AND B.~J. {Drouin} AND J.-M. {Flaud} AND R.~R. {Gamache} AND J.~T. {Hodges} AND D. {Jacquemart} AND E.~J. {Mlawer} AND A.~V. {Nikitin} AND V.~I. {Perevalov} AND M. {Rotger} AND J. {Tennyson} AND G.~C. {Toon} AND H. {Tran} AND V.~G. {Tyuterev} AND E.~M. {Adkins} AND A. {Baker} AND A. {Barbe} AND E. {Can{\`{e}}} AND A.~G. {Cs{'{a}}sz{'{a}}r} AND A. {Dudaryonok} AND O. {Egorov} AND A.~J. {Fleisher} AND H. {Fleurbaey} AND A. {Foltynowicz} AND T. {Furtenbacher} AND J.~J. {Harrison} AND J.-M. {Hartmann} AND V.-M. {Horneman} AND X. {Huang} AND T. {Karman} AND J. {Karns} AND S. {Kassi} AND I. {Kleiner} AND V. {Kofman} AND F. {Kwabia-Tchana} AND N.~N. {Lavrentieva} AND T.~J. {Lee} AND D.~A. {Long} AND A.~A. {Lukashevskaya} AND O.~M. {Lyulin} AND V.~Yu. {Makhnev} AND W. {Matt} AND S.~T. {Massie} AND M. {Melosso} AND S.~N. {Mikhailenko} AND D. {Mondelain} AND H.~S.~P. {M{"{u}}ller} AND O.~V. {Naumenko} AND A. {Perrin} AND O.~L. {Polyansky} AND E. {Raddaoui} AND P.~L. {Raston} AND Z.~D. {Reed} AND M. {Rey} AND C. {Richard} AND R. {T{'{o}}bi{'{a}}s} AND I. {Sadiek} AND D.~W. {Schwenke} AND E. {Starikova} AND K. {Sung} AND F. {Tamassia} AND S.~A. {Tashkun} AND J. {Vander Auwera} AND I.~A. {Vasilenko} AND A.~A. {Vigasin} AND G.~L. {Villanueva} AND B. {Vispoel} AND G. {Wagner} AND A. {Yachmenev} AND S.~N. {Yurchenko}},
title = {The {HITRAN2020} Molecular Spectroscopic Database},
journal = {Journal of Quantitative Spectroscopy and Radiative Transfer},
year = {2022},
volume = {277},
pages = {107949},
doi = {10.1016/j.jqsrt.2021.107949},
}

@article{Kochanov2019,
author = {{Kochanov}, R.V and {Gordon}, I. E and {Rothman}, L. S. and {Shine}, K. P. and {Sharpe}, S. W. and {Johnson}, T. J. and {Wallington}, T. J. and {Harrison}, J. J. and {Bernath}, P. F. and {Birk}, M. and {Wagner}, G. and {Le Bris}, K. and {Bravo}, I. and {Hill},  C.},
title = {Infrared absorption cross-sections in HITRAN2016 and beyond: Expansion for climate, environment, and atmospheric applications},
journal = {Journal of Quantitative Spectroscopy and Radiative Transfer},
year = {2019},
doi = {10.1016/j.jqsrt.2019.04.001},
}

@article{Sharpe2004,
author = {S. W. Sharpe AND  T. J. Johnson AND  R. L. Sams AND  P. M. Chu AND  G. C. Rhoderick AND  P. A. Johnson},
title = {Gas-Phase Databases for Quantitative Infrared Spectroscopy},
journal = {Applied Spectroscopy},
year = {2004},
volume = {58},
pages = {1452-1461},
doi = {10.1366/0003702042641281},
}

@ARTICLE{Haser1957,
       author = {{Haser}, L.},
        title = "{Distribution d'intensit{\'e} dans la t{\^e}te d'une com{\`e}te}",
      journal = {Bulletin de la Societe Royale des Sciences de Liege},
         year = 1957,
        month = jan,
       volume = {43},
        pages = {740-750},
       adsurl = {https://ui.adsabs.harvard.edu/abs/1957BSRSL..43..740H}
}

@ARTICLE{Villanueva2012b,
       author = {{Villanueva}, G.~L. and {DiSanti}, M.~A. and {Mumma}, M.~J. and {Xu}, L. -H.},
        title = "{A Quantum Band Model of the {\ensuremath{\nu}}$_{3}$ Fundamental of Methanol (CH$_{3}$OH) and Its Application to Fluorescence Spectra of Comets}",
      journal = {Astrophysical Journal},
         year = 2012,
        month = mar,
       volume = {747},
       number = {1},
          eid = {37},
        pages = {37},
          doi = {10.1088/0004-637X/747/1/37},
       adsurl = {https://ui.adsabs.harvard.edu/abs/2012ApJ...747...37V}
}

@ARTICLE{Villanueva2012a,
       author = {{Villanueva}, G.~L. and {Mumma}, M.~J. and {Bonev}, B.~P. and {Novak}, R.~E. and {Barber}, R.~J. and {Disanti}, M.~A.},
        title = "{Water in planetary and cometary atmospheres: H$_{2}$O/HDO transmittance and fluorescence models}",
      journal = {Journal of Quantitative Spectroscopy and Radiative Transfer},
         year = 2012,
        month = feb,
       volume = {113},
       number = {3},
        pages = {202-220},
          doi = {10.1016/j.jqsrt.2011.11.001},
       adsurl = {https://ui.adsabs.harvard.edu/abs/2012JQSRT.113..202V}
}

@article{DelloRusso2000,
title = {Water Production and Release in Comet C/1995 O1 Hale–Bopp},
journal = {Icarus},
volume = {143},
number = {2},
pages = {324-337},
year = {2000},
issn = {0019-1035},
doi = {https://doi.org/10.1006/icar.1999.6268},
url = {https://www.sciencedirect.com/science/article/pii/S0019103599962681},
author = {Neil {Dello Russo} and Michael J. Mumma and Michael A. DiSanti and Karen Magee-Sauer and Robert Novak and Terrence W. Rettig}
}

@article{Tubergen2000,
author = {Tubergen, Michael J. and Lavrich, Richard J. and McCargar, James W.},
title = {Infrared Spectrum and Group Theoretical Analysis of the Vibrational Modes of Carbonyl Sulfide},
journal = {Journal of Chemical Education},
volume = {77},
number = {12},
pages = {1637},
year = {2000},
doi = {10.1021/ed077p1637},
URL = {https://doi.org/10.1021/ed077p1637},
eprint = {https://doi.org/10.1021/ed077p1637}
}

@ARTICLE{Villanueva2011,
       author = {{Villanueva}, G.~L. and {Mumma}, M.~J. and {DiSanti}, M.~A. and {Bonev}, B.~P. and {Gibb}, E.~L. and {Magee-Sauer}, K. and {Blake}, G.~A. and {Salyk}, C.},
        title = "{The molecular composition of Comet C/2007 W1 (Boattini): Evidence of a peculiar outgassing and a rich chemistry}",
      journal = {Icarus},
         year = 2011,
        month = nov,
       volume = {216},
       number = {1},
        pages = {227-240},
          doi = {10.1016/j.icarus.2011.08.024},
       adsurl = {https://ui.adsabs.harvard.edu/abs/2011Icar..216..227V}
}

@article{Horka2004,
author = {Horká, Veronika and Civiš, Svatopluk and Špirko, Vladim\'ir and Kawaguchi, Kentarou},
year = {2004},
month = {01},
pages = {73--89},
title = {The Infrared Spectrum of CN in Its Ground Electronic State},
volume = {69},
journal = {Collection of Czechoslovak Chemical Communications},
doi = {10.1135/cccc20040073}
}

@article{Ellis1931,
  title = {Polymers and New Infrared Absorption Bands of Water},
  author = {Ellis, Joseph W.},
  journal = {Physical Review},
  volume = {38},
  issue = {4},
  pages = {693--698},
  numpages = {0},
  year = {1931},
  month = {Aug},
  publisher = {American Physical Society},
  doi = {10.1103/PhysRev.38.693},
  url = {https://link.aps.org/doi/10.1103/PhysRev.38.693}
}

@ARTICLE{Cleeves2014,
       author = {{Cleeves}, L. Ilsedore and {Bergin}, Edwin A. and {Alexander}, Conel M.~O. 'D. and {Du}, Fujun and {Graninger}, Dawn and {{\"O}berg}, Karin I. and {Harries}, Tim J.},
        title = "{The ancient heritage of water ice in the solar system}",
      journal = {Science},
         year = 2014,
        month = sep,
       volume = {345},
       number = {6204},
        pages = {1590-1593},
          doi = {10.1126/science.1258055},
archivePrefix = {arXiv},
       eprint = {1409.7398},
 primaryClass = {astro-ph.SR},
       adsurl = {https://ui.adsabs.harvard.edu/abs/2014Sci...345.1590C}
}

@article{Zeichner2023,
author = {Sarah S. Zeichner  and José C. Aponte  and Surjyendu Bhattacharjee  and Guannan Dong  and Amy E. Hofmann  and Jason P. Dworkin  and Daniel P. Glavin  and Jamie E. Elsila  and Heather V. Graham  and Hiroshi Naraoka  and Yoshinori Takano  and Shogo Tachibana  and Allison T. Karp  and Kliti Grice  and Alex I. Holman  and Katherine H. Freeman  and Hisayoshi Yurimoto  and Tomoki Nakamura  and Takaaki Noguchi  and Ryuji Okazaki  and Hikaru Yabuta  and Kanako Sakamoto  and Toru Yada  and Masahiro Nishimura  and Aiko Nakato  and Akiko Miyazaki  and Kasumi Yogata  and Masanao Abe  and Tatsuaki Okada  and Tomohiro Usui  and Makoto Yoshikawa  and Takanao Saiki  and Satoshi Tanaka  and Fuyuto Terui  and Satoru Nakazawa  and Sei-ichiro Watanabe  and Yuichi Tsuda  and Kenji Hamase  and Kazuhiko Fukushima  and Dan Aoki  and Minako Hashiguchi  and Hajime Mita  and Yoshito Chikaraishi  and Naohiko Ohkouchi  and Nanako O. Ogawa  and Saburo Sakai  and Eric T. Parker  and Hannah L. McLain  and Francois-Regis Orthous-Daunay  and Véronique Vuitton  and Cédric Wolters  and Philippe Schmitt-Kopplin  and Norbert Hertkorn  and Roland Thissen  and Alexander Ruf  and Junko Isa  and Yasuhiro Oba  and Toshiki Koga  and Toshihiro Yoshimura  and Daisuke Araoka  and Haruna Sugahara  and Aogu Furusho  and Yoshihiro Furukawa  and Junken Aoki  and Kuniyuki Kano  and Shin-ichiro M. Nomura  and Kazunori Sasaki  and Hajime Sato  and Takaaki Yoshikawa  and Satoru Tanaka  and Mayu Morita  and Morihiko Onose  and Fumie Kabashima  and Kosuke Fujishima  and Tomoya Yamazaki  and Yuki Kimura  and John M. Eiler },
title = {Polycyclic aromatic hydrocarbons in samples of Ryugu formed in the interstellar medium},
journal = {Science},
volume = {382},
number = {6677},
pages = {1411-1416},
year = {2023},
doi = {10.1126/science.adg6304},
URL = {https://www.science.org/doi/abs/10.1126/science.adg6304},
eprint = {https://www.science.org/doi/pdf/10.1126/science.adg6304}}

@misc{jwst,
    author = {{Bushouse}, Howard and {Eisenhamer}, Johnathan and {Dencheva}, Nadia and {Davies}, James and {Greenfield}, Perry and {Morrison}, Jane and {Hodge}, Phil and {Simon}, Bernie and {Grumm}, David and {Droettboom}, Michael and {Slavich}, Edward and {Sosey}, Megan and {Pauly}, Tyler and {Miller}, Todd and {Jedzejewski}, Robert and {Hack}, Warren and {Davis}, David and {Crawford}, Stephen and {Law}, David and {Gordon}, Karl and {Reagan}, Michael and {Cara}, Mihai and {MacDonald}, Ken and {Bradley}, Larry and {Shanahan}, Clare and {Jamieson}, William and {Teodoro}, Marian and {Williams}, Thomas and {Pena-Guerrero}, Maria and {Graham}, Brett and {Molter}, Edward and {Brandt}, Timothy and {Hayes}, Christian and {Cooper}, Rachel and {Clarke}, Melanie and {Filippazzo}, Joseph},
    title = {JWST Calibration Pipeline},
    doi = {10.5281/zenodo.7038885},
    year = 2024,
    publisher = {Zenodo},
    version = {1.16.0},
    url = {https://github.com/spacetelescope/jwst}
}

@ARTICLE{Rodgers2002,
       author = {{Rodgers}, S.~D. and {Charnley}, S.~B.},
        title = "{A model of the chemistry in cometary comae: deuterated molecules}",
      journal = {Monthly Notices of the Royal Astronomical Society},
         year = 2002,
        month = mar,
       volume = {330},
       number = {3},
        pages = {660-674},
          doi = {10.1046/j.1365-8711.2002.05165.x},
       adsurl = {https://ui.adsabs.harvard.edu/abs/2002MNRAS.330..660R}
}

@ARTICLE{Oberg2023,
       author = {{{\"O}berg}, Karin I. and {Facchini}, Stefano and {Anderson}, Dana E.},
        title = "{Protoplanetary Disk Chemistry}",
      journal = {Annual Review of Astronomy and Astrophysics},
         year = 2023,
        month = aug,
       volume = {61},
        pages = {287-328},
          doi = {10.1146/annurev-astro-022823-040820},
archivePrefix = {arXiv},
       eprint = {2309.05685},
 primaryClass = {astro-ph.EP},
       adsurl = {https://ui.adsabs.harvard.edu/abs/2023ARA\&A..61..287O}
}

@article{casa,
    author = {{The CASA Team} and {Bean}, Ben and {Bhatnagar}, Sanjay and {Castro}, Sandra and {Meyer}, Jennifer Donovan and {Emonts}, Bjorn and {Garcia}, Enrique and {Garwood}, Robert and {Golap}, Kumar and {Villalba}, Justo Gonzalez and {Harris}, Pamela and {Hayashi}, Yohei and {Hoskins}, Josh and {Hseih}, Mingyu and {Jagannathan}, Preshanth and  {Kawasaki}, Wataru and {Keimpema}, Aard and {Kettenis}, Mark and {Lopez}, Jorge and {Marvil}, Joshua and {Masters}, Joseph anf {McNichols}, Andrew and {Mehringer}, David and {Miel}, Renaud andj {Moellenbrock}, George and {Montesino}, Federico and {Nakazato}, Takeshi and {Ott}, Juergen and {Petry}, Dirk and {Pokorny}, Martin and {Raba}, Ryan and {Rau}, Urvashi and {Schiebel}, Darrell and {Schweighart}, Neal and {Sekhar}, Srikrishna and {Shimada}, Kazuhiko and {Small}, Des and {Steeb}, Jan-Willem and {Sugimoto}, Kanako and {Suoranta}, Ville and {Tsutsumi}, Takahiro and {van Bemmel}, Ilse M. and {Vekouter}, Marjolein and {Wells}, Akeem and {Xiong}, Wei and {Szomoru}, Arpad and {Griffith}, Morgan and {Glendenning}, Brian and {Kern}, Jeff},
    title = {CASA, the Common Astronomy Software Applications for Radio Astronomy},
    journal = {Publications of the Astronomical Society of the Pacific},
    year = 2022,
    volume = 134,
    number = 1041,
    doi = {10.1088/1538-3873/ac9642}
}

@misc{jdaviz,
  author       = {{JDADF Developers} and
                  Averbukh, Jesse and
                  Bradley, Larry and
                  Buikhuizen, Mario and
                  Busko, Ivo and
                  Cherinka, Brian and
                  Conroy, Kyle and
                  Earl, Nicholas and
                  Fox, Ori and
                  Geda, Robel and
                  Green, Gilbert and
                  Jones, Craig and
                  Karatay, Hatice and
                  Kotler, Jenn and
                  Lim, Pey Lian and
                  Morris, Brett and
                  Nguyen, Duy and
                  O'Steen, Richard and
                  Ogaz, Sara and
                  Ogle, Patrick and
                  Otor, O. Justin and
                  Pacifici, Camilla and
                  Robitaille, Thomas and
                  Shanahan, Clare and
                  Tollerud, Erik and
                  Volfman, Sabrina},
  title        = {Jdaviz},
  month        = oct,
  year         = 2024,
  publisher    = {Zenodo},
  version      = {v3.10.4},
  doi          = {10.5281/zenodo.14009504},
  url          = {https://doi.org/10.5281/zenodo.14009504}
}

@book{Shimanouchi1972,
    author = {{Shimanouchi}, Takehiko},
    title = {Tables of Molecular Vibrational Frequencies, Consolidated Volume I},
    series = {NBS National Standard Reference Data Series},
    number = {39},
    publisher = {National Bureau of Standards},
    year = 1972
}

@article{Bonner1934,
  title = {The Vibrational Spectrum of Water Vapor},
  author = {Bonner, Lyman G},
  journal = {Physical Review},
  volume = {46},
  number = {6},
  pages = {458--464},
  year = {1934},
  month = {Sep},
  publisher = {American Physical Society},
  doi = {10.1103/PhysRev.46.458},
  url = {https://link.aps.org/doi/10.1103/PhysRev.46.458}
}

@article{Endres2016,
title = {The Cologne Database for Molecular Spectroscopy, CDMS, in the Virtual Atomic and Molecular Data Centre, VAMDC},
journal = {Journal of Molecular Spectroscopy},
volume = {327},
pages = {95-104},
year = {2016},
note = {New Visions of Spectroscopic Databases, Volume II},
issn = {0022-2852},
doi = {https://doi.org/10.1016/j.jms.2016.03.005},
url = {https://www.sciencedirect.com/science/article/pii/S0022285216300340},
author = {Christian P. Endres and Stephan Schlemmer and Peter Schilke and Jürgen Stutzki and Holger S.P. Müller}
}

@ARTICLE{Manzini2023,
       author = {{Manzini}, Federico and {Oldani}, Virginio and {Ochner}, Paolo and {Bedin}, Luigi R. and {Reguitti}, Andrea},
        title = "{Rotation period and Morphological Structures in the inner coma of comet C/2022 E3 (ZTF)}",
      journal = {The Astronomer's Telegram},
         year = 2023,
        month = feb,
       volume = {15909},
        pages = {1},
       adsurl = {https://ui.adsabs.harvard.edu/abs/2023ATel15909....1M}
}

@ARTICLE{Knight2023,
       author = {{Knight}, M.~M. and {Holt}, C.~E. and {Villa}, K.~M. and {Skiff}, B.~A. and {Schleicher}, D.~G.},
        title = "{Rotation period of comet C/2022 E3 ZTF from CN morphology}",
      journal = {The Astronomer's Telegram},
         year = 2023,
        month = jan,
       volume = {15879},
        pages = {1},
       adsurl = {https://ui.adsabs.harvard.edu/abs/2023ATel15879....1K}
}

@ARTICLE{Liu2024,
       author = {{Liu}, Bin and {Liu}, Xiaodong},
        title = "{Unraveling the dust activity of naked-eye comet C/2022 E3 (ZTF)}",
      journal = {Astronomy and Astrophysics},
         year = 2024,
        month = mar,
       volume = {683},
          eid = {A51},
        pages = {A51},
          doi = {10.1051/0004-6361/202348663},
archivePrefix = {arXiv},
       eprint = {2405.15351},
 primaryClass = {astro-ph.EP},
       adsurl = {https://ui.adsabs.harvard.edu/abs/2024A\&A...683A..51L}
}

@incollection{Meech2004,
    author = {{Meech}, K. J. and {Svore\v{n}}, J.},
    title = {Using Cometary Activity to Trace the Physical and Chemical Evolution of Cometary Nuclei},
    pages = {317--335},
    booktitle = {Comets II},
    editor = {{Festou}, M. C. and {Keller}, H. U. and {Weaver}, H. A.},
    publisher = {The University of Arizona Press},
    address = {Tuscon, AZ},
    year = 2004
}

@incollection{Rodgers2004,
    author = {{Rodgers}, S. D. and {Charnley}, S. B. and {Huebner}, W. F. and {Boice}, D. C.},
    title = {Physical Processes and Chemical Reactions in Cometary Comae},
    pages = {505--522},
    booktitle = {Comets II},
    editor = {{Festou}, M. C. and {Keller}, H. U. and {Weaver}, H. A.},
    publisher = {The University of Arizona Press},
    address = {Tuscon, AZ},
    year = 2004
}

@ARTICLE{Weissman2020,
       author = {{Weissman}, Paul and {Morbidelli}, Alessandro and {Davidsson}, Bj{\"o}rn and {Blum}, J{\"u}rgen},
        title = "{Origin and Evolution of Cometary Nuclei}",
      journal = {Space Science Reviews},
         year = 2020,
        month = jan,
       volume = {216},
       number = {1},
          eid = {6},
        pages = {6},
          doi = {10.1007/s11214-019-0625-7},
       adsurl = {https://ui.adsabs.harvard.edu/abs/2020SSRv..216....6W}
}

@ARTICLE{Dones2015,
       author = {{Dones}, Luke and {Brasser}, Ramon and {Kaib}, Nathan and {Rickman}, Hans},
        title = "{Origin and Evolution of the Cometary Reservoirs}",
      journal = {Space Science Reviews},
         year = 2015,
        month = dec,
       volume = {197},
       number = {1-4},
        pages = {191-269},
          doi = {10.1007/s11214-015-0223-2},
       adsurl = {https://ui.adsabs.harvard.edu/abs/2015SSRv..197..191D}
}

@ARTICLE{Bolin2022,
       author = {{Bolin}, B.~T. and {Masci}, F.~J. and {Ip}, W. -H. and {Helou}, G. and {Kramer}, E.~A. and {Lin}, Z. -Y. and {Prince}, T.~A. and {Sato}, H. and {Paul}, N. and {Yoshimoto}, K. and {Urbanik}, M. and {Denneau}, L. and {Siverd}, R. and {Tonry}, J. and {Weiland}, H. and {Erasmus}, N. and {Fitzsimmons}, A. and {Lawrence}, A. and {Robinson}, J. and {Siverd}, R. and {Tonry}, J. and {Birtwhistle}, P. and {Jacques}, C. and {Hug}, G. and {Korlevic}, K. and {Buzzi}, L. and {Bacci}, R. and {van Buitenen}, G. and {Buczynski}, D. and {Hale}, A. and {Masek}, M. and {Guido}, E. and {Rocchetto}, M. and {Bryssinck}, E. and {Milani}, G. and {Savini}, G. and {Valvasori}, A. and {Ligustri}, R. and {Bacci}, P. and {Maestripieri}, M. and {Tesi}, L. and {Fagioli}, G. and {Lutkenhoner}, B.},
        title = "{Comet C/2022 E3 (ZTF)}",
      journal = {Minor Planet Electronic Circulars},
         year = 2022,
        month = mar,
       volume = {2022-F13},
       adsurl = {https://ui.adsabs.harvard.edu/abs/2022MPEC....F...13B}
}

@ARTICLE{Lippi2024,
       author = {{Lippi}, Manuela and {Podio}, Linda and {Codella}, Claudio and {Faggi}, Sara and {De Simone}, Marta and {Villanueva}, Geronimo L. and {Mumma}, Michael J. and {Ceccarelli}, Cecilia},
        title = "{The Ice Chemistry in Comets and Planet-forming Disks: Statistical Comparison of CH$_{3}$OH, H$_{2}$CO, and NH$_{3}$ Abundance Ratios}",
      journal = {Astrophysical Journal, Letters},
         year = 2024,
        month = jul,
       volume = {970},
       number = {1},
          eid = {L5},
        pages = {L5},
          doi = {10.3847/2041-8213/ad5a6d},
       adsurl = {https://ui.adsabs.harvard.edu/abs/2024ApJ...970L...5L}
}

@ARTICLE{Lippi2021,
       author = {{Lippi}, M. and {Villanueva}, G.~L. and {Mumma}, M.~J. and {Faggi}, S.},
        title = "{Investigation of the Origins of Comets as Revealed through Infrared High-resolution Spectroscopy I. Molecular Abundances}",
      journal = {Astronomical Journal},
         year = 2021,
        month = aug,
       volume = {162},
       number = {2},
          eid = {74},
        pages = {74},
          doi = {10.3847/1538-3881/abfdb7},
       adsurl = {https://ui.adsabs.harvard.edu/abs/2021AJ....162...74L}
}

@ARTICLE{Biver2024a,
       author = {{Biver}, N. and {Bockel{\'e}e-Morvan}, D. and {Handzlik}, B. and {Sandqvist}, Aa. and {Boissier}, J. and {Drozdovskaya}, M.~N. and {Moreno}, R. and {Crovisier}, J. and {Lis}, D.~C. and {Cordiner}, M. and {Milam}, S. and {Roth}, N.~X. and {Bonev}, B.~P. and {Dello Russo}, N. and {Vervack}, R. and {Opitom}, C. and {Kawakita}, H.},
        title = "{Chemical composition of comets C/2021 A1 (Leonard) and C/2022 E3 (ZTF) from radio spectroscopy and the abundance of HCOOH and HNCO in comets}",
      journal = {Astronomy and Astrophysics},
         year = 2024,
        month = oct,
       volume = {690},
          eid = {A271},
        pages = {A271},
          doi = {10.1051/0004-6361/202450921},
archivePrefix = {arXiv},
       eprint = {2408.10759},
 primaryClass = {astro-ph.EP},
       adsurl = {https://ui.adsabs.harvard.edu/abs/2024A\&A...690A.271B}
}

@article{Czekala2021,
    author = {{Czekala}, Ian and {Loomis}, Ryan A. and {Teague}, Richard and {Booth}, Alice S. and {Huang}, Jane and {Cataldi}, Gianni and {Ilee}, John D. and {Law}, Charles J. and {Walsh}, Catherine and {Bosman}, Arthur D. and {Guzm\'an}, Viviana V. and {Le Gal}, Romane and {\"Oberg}, Karin I. and {Yamato}, Yoshihide and {Aikawa}, Yuri and {Andrew}, Sean M. and {Bae}, Jaehan and {Bergin}, Edwin A. and {Bergner}, Jennifer B. and {Cleeves}, L. Ilsedore and {Kurtovic}< Nicholas T. and {M\'enard}, Fran\c cois and {Nomura}, Hideko and {P\'erez}, Laura M. and {Qi}, Chunhua and {Schwarz}, Kamber R. and {Tsukagoshi}, Takashi and {Waggoner}, Abygail R. and {Wilner}, David J. and {Zhang}, Ke},
    title = {Molecules with ALMA at Planet-forming Scales (MAPS). II. CLEAN Strategies for Synthesizing Images of Molecular Line Emission in Protoplanetary Disks},
    journal = {The Astrophysical Journal Supplement Series},
    year = 2021,
    month = nov,
    volume = {257},
    number = {2},
    pages = {19p},
    doi = {10.3847/1538-4365/ac1430}
}

@article{Cordiner2017,
       author = {{Cordiner}, M.~A. and {Biver}, N. and {Crovisier}, J. and {Bockel{\'e}e-Morvan}, D. and {Mumma}, M.~J. and {Charnley}, S.~B. and {Villanueva}, G. and {Paganini}, L. and {Lis}, D.~C. and {Milam}, S.~N. and {Remijan}, A.~J. and {Coulson}, I.~M. and {Kuan}, Y-J. and {Boissier}, J.},
        title = "{Thermal Physics of the Inner Coma: ALMA Studies of the Methanol Distribution and Excitation in Comet C/2012 K1 (PanSTARRS)}",
      journal = {The Astronomical Journal},
         year = 2017,
        month = mar,
       volume = {837},
       number = {2},
          eid = {177},
        pages = {177},
          doi = {10.3847/1538-4357/aa6211},
archivePrefix = {arXiv},
       eprint = {1701.08258},
 primaryClass = {astro-ph.EP},
       adsurl = {https://ui.adsabs.harvard.edu/abs/2017ApJ...837..177C}
}

@inproceedings{Villanueva2017,
       author = {{Villanueva}, G.~L. and {Mandell}, A. and {Protopapa}, S. and {Faggi}, S. and {Smith}, M.~D. and {Wolff}, M. and {Hewagama}, T. and {Mumma}, M.~J.},
        title = "{Planetary Spectrum Generator (PSG): An Online Tool to Synthesize Spectra of Comets, Small Bodies, and (Exo)Planets}",
    booktitle = {Planetary Science Vision 2050 Workshop},
         year = 2017,
       editor = {{LPI Editorial Board}},
       series = {LPI Contributions},
       volume = {1989},
        month = feb,
          eid = {8006},
        pages = {8006},
       adsurl = {https://ui.adsabs.harvard.edu/abs/2017LPICo1989.8006V}
}

@article{Cordiner2023,
   author = {M. A. Cordiner and N. X. Roth and S. N. Milam and G. L. Villanueva and D. Bockelée-Morvan and A. J. Remijan and S. B. Charnley and N. Biver and D. C. Lis and C. Qi and B. P. Bonev and J. Crovisier and J. Boissier},
   doi = {10.3847/1538-4357/ace0bc},
   issn = {0004-637X},
   issue = {1},
   journal = {The Astrophysical Journal},
   month = {8},
   pages = {59},
   title = {Gas Sources from the Coma and Nucleus of Comet 46P/Wirtanen Observed Using ALMA},
   volume = {953},
   url = {https://iopscience.iop.org/article/10.3847/1538-4357/ace0bc},
   year = {2023},
}

@article{Mumma2011,
   author = {Michael J. Mumma and Steven B. Charnley},
   doi = {10.1146/annurev-astro-081309-130811},
   issn = {00664146},
   journal = {Annual Review of Astronomy and Astrophysics},
   month = {9},
   pages = {471-524},
   publisher = {Annual Reviews Inc.},
   title = {The chemical composition of cometsemerging taxonomies and natal heritage},
   volume = {49},
   year = {2011},
}

@incollection{Biver2024b,
    author = {{Biver}, N and Neil Dello Russo and Cyrielle Opitom and Martin Rubin},
    title = {Chemistry of Comet Atmospheres},
    booktitle = {Comets III},
    editor = {{Meech}, Karen J. and {Combi}, Michael R. and {Bockl\'ee-Morvan}, Dominique and {Raymond}, Sean N. and Michael Zolensky},
    publisher = {University of Arizona Press},
    isbn = {9780816553631},
    year = {2024},
    month = {12},
    pages = {459--498}
}

@article{Drozdovskaya2019,
   author = {Maria N. Drozdovskaya and Ewine F. Van Dishoeck and Martin Rubin and Jes K. Jørgensen and Kathrin Altwegg},
   doi = {10.1093/mnras/stz2430},
   issn = {13652966},
   issue = {1},
   journal = {Monthly Notices of the Royal Astronomical Society},
   month = {11},
   pages = {50-79},
   publisher = {Oxford University Press},
   title = {Ingredients for solar-like systems: Protostar IRAS 16293-2422 B versus comet 67P/Churyumov–Gerasimenko},
   volume = {490},
   year = {2019},
}

@article{Caselli2012,
   author = {Paola Caselli and Cecilia Ceccarelli},
   issn = {09354956},
   issue = {1},
   journal = {Astronomy and Astrophysics Review},
   month = {10},
   publisher = {Springer Verlag},
   title = {Our astrochemical heritage},
   volume = {20},
   year = {2012},
}

@article{Bolin2024a,
   author = {B. T. Bolin and F. J. Masci and D. A. Duev and J. W. Milburn and J. D. Neill and J. N. Purdum and C. Avdellidou and M. Saki and Y. C. Cheng and M. Delbo and C. Fremling and M. Ghosal and Z. Y. Lin and C. M. Lisse and A. Mahabal},
   doi = {10.1093/mnrasl/slad139},
   issn = {17453933},
   issue = {1},
   journal = {Monthly Notices of the Royal Astronomical Society: Letters},
   month = {1},
   pages = {L42-L46},
   publisher = {Oxford University Press},
   title = {Palomar discovery and initial characterization of naked-eye long-period comet C/2022 E3 (ZTF)},
   volume = {527},
   year = {2024},
}
\bibliographystyle{aasjournalv7}

\newpage
\appendix
\counterwithin{figure}{section}

\section{SUPPLEMENTAL
FIGURES} \label{appendix}

\begin{figure}
    \centering
    \includegraphics[width=\linewidth]{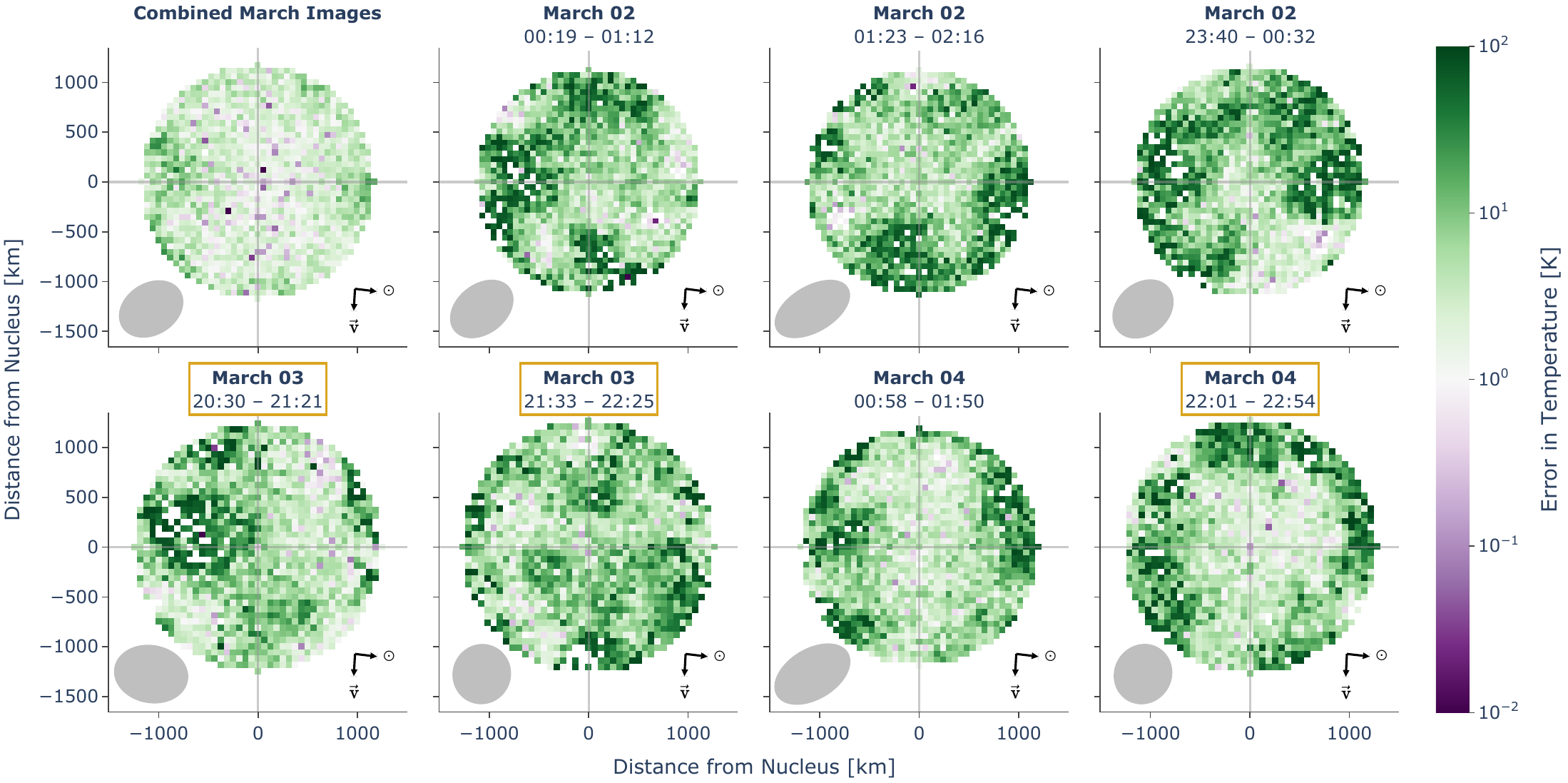}
    \caption{Maps depicting error in the modeled temperature maps shown in Figure \ref{fig_ATmaps}. Observations that were measured during the day or shortly after sunset are indicated by yellow label borders. Arrows in the bottom-right corner indicate the direction of the Sun ($\odot$) and comet velocity ($\vec{\mathrm{v}}$). The ALMA synthesized  beam for each image is shown in the bottom-left corner.}
    \label{fig_ATerrmaps}
\end{figure}

\begin{figure}
    \centering
    \includegraphics[width=\linewidth]{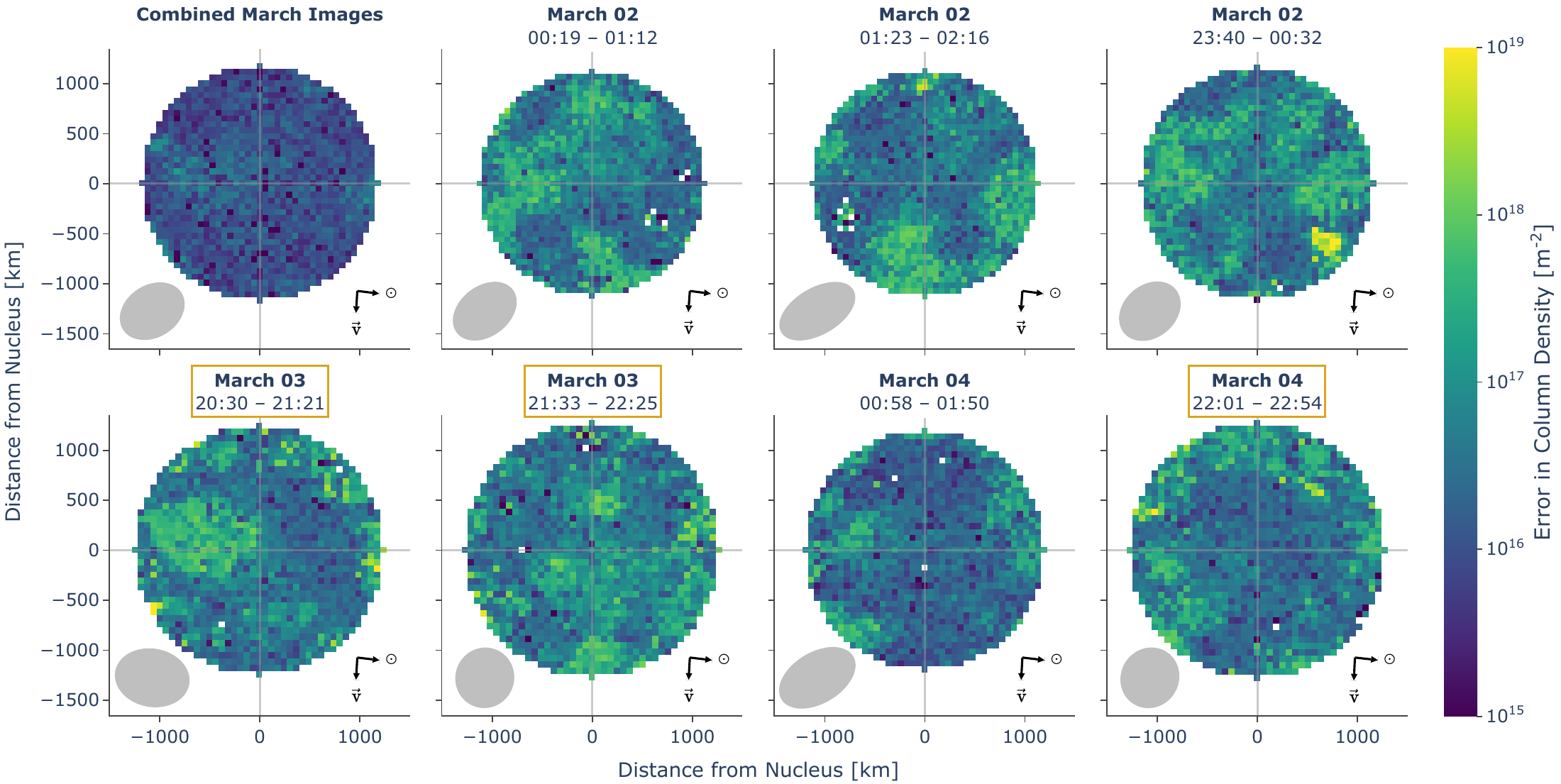}
    \caption{Maps depicting error in the modeled column density maps shown in Figure \ref{fig_ANmaps}. Observations that were measured during the day or shortly after sunset are indicated by yellow label borders. Arrows in the bottom-right corner indicate the direction of the Sun ($\odot$) and comet velocity ($\vec{\mathrm{v}}$). The ALMA synthesized  beam for each image is shown in the bottom-left corner.}
    \label{fig_ANerrmaps}
\end{figure}

\begin{figure*}
    \centering
    \includegraphics[width=\linewidth]{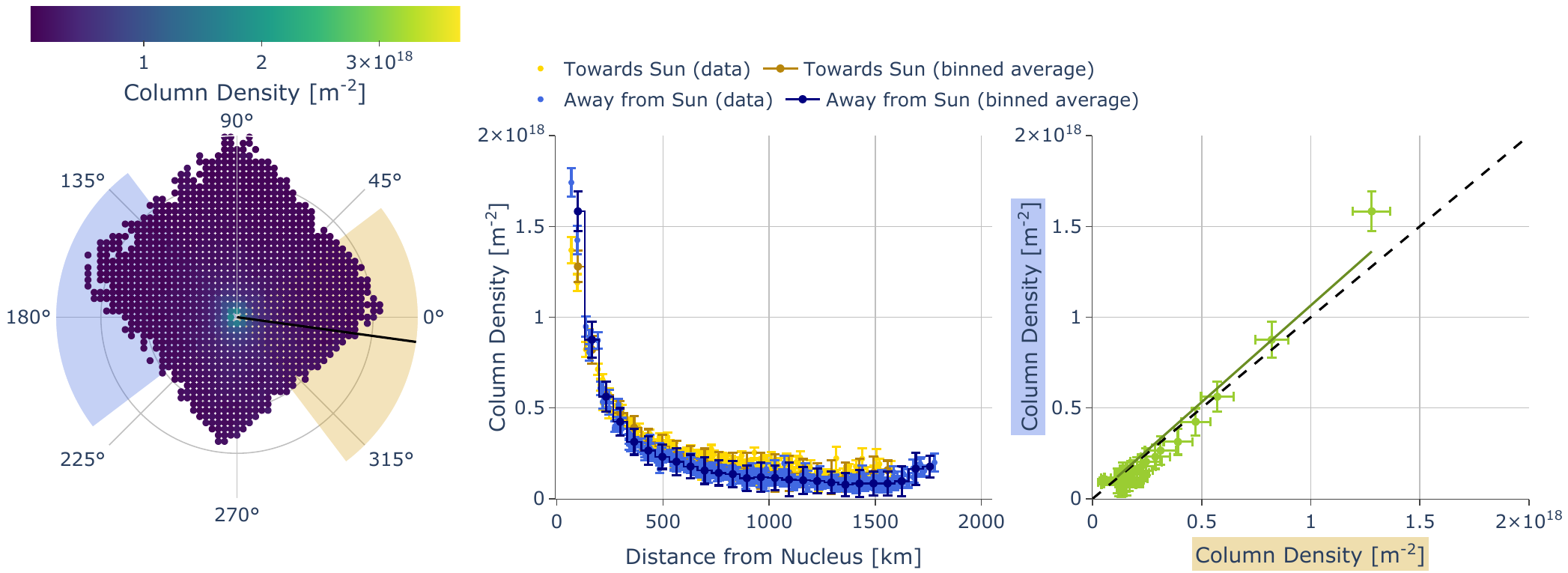}
    \caption{(left)  Modeled \ce{CH3OH} column density map for JWST region [d] in Figure \ref{fig_Jspec} with regions within ± 45º of both the Sunward (yellow) and anti-Sunward (blue) vectors. (center) Distributions of the modeled \ce{CH3OH} column density for the Sun- and anti-Sun regions versus radial distance with bin-averaged distributions overlaid. (right) Bin-averaged modeled \ce{CH3OH} column density for the away from Sun region (blue) against the towards Sun region (yellow). The linear fit has an equation of $y = (1.064 \pm 0.040) \cdot x + (1.965 \pm 0.073) \times 10^{-18}$ and an $\mathrm{R}^2 = 0.9512$ .}
    \label{fig_antisun-sun-Nch3oh}
\end{figure*}

\end{document}